\documentstyle[eqsecnum,aps,psfig]{revtex}
\begin{document}
\draft
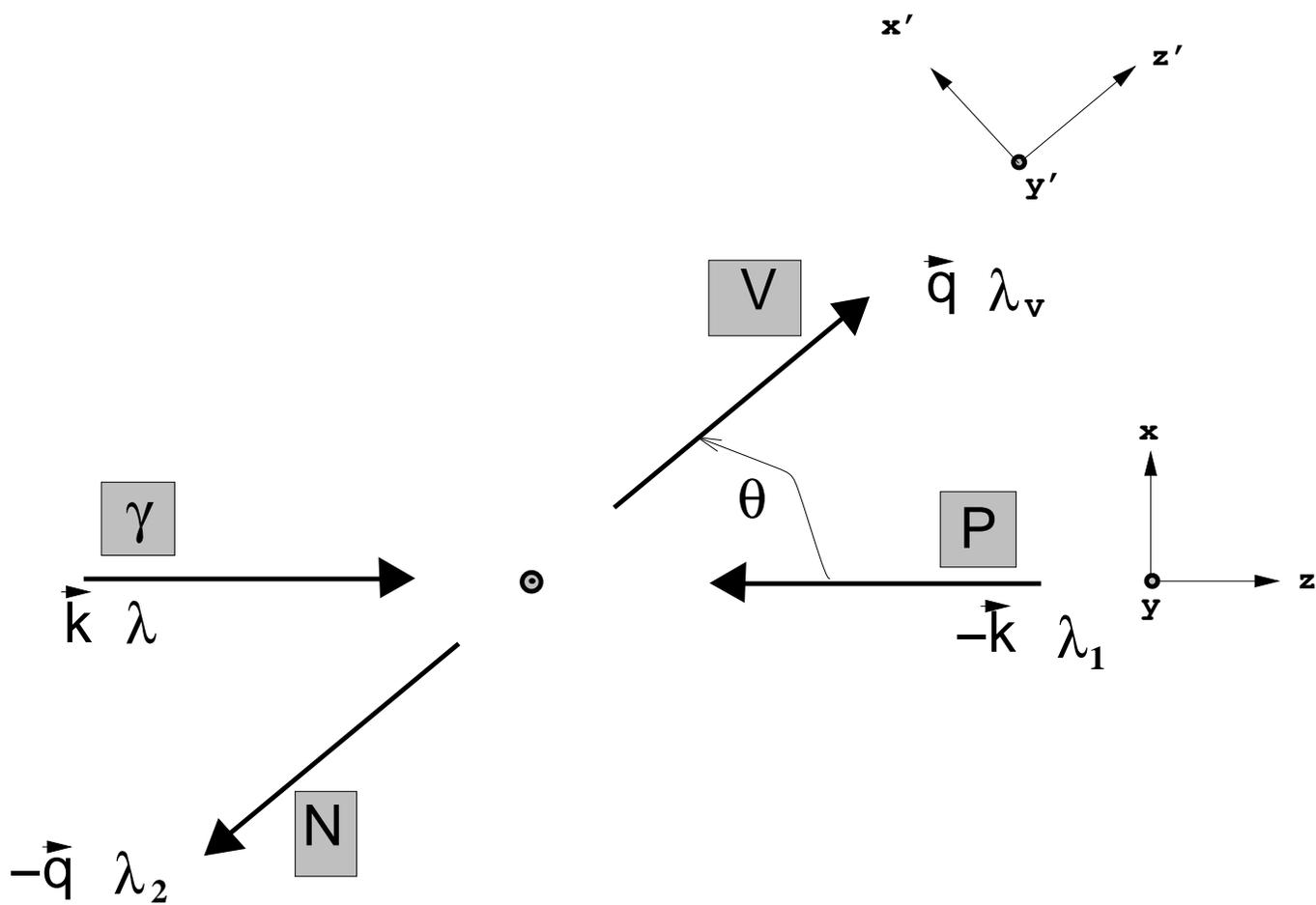
\title{ Polarization observables in vector meson photoproduction}
\author{Michael Pichowsky, \c{C}etin \c{S}avkl\i\ and Frank
Tabakin~\cite{byline} }
\address{ Department of Physics \& Astronomy, University of
Pittsburgh, Pittsburgh, Pennsylvania  15260}
\date{\today}
\maketitle
\begin{abstract}
The photoproduction of vector mesons($\rho, \omega, \phi$)
 is of renewed interest because
 intense  high energy beams of
 polarized electrons and photons are under development.
   These beams and also polarized targets
make it possible to explore the dynamics of basic baryon structure.
As a step toward  that goal, an analysis of all possible polarization
observables for the case of vector meson photoproduction from a
nucleon
target is presented.
 The question of which observables are needed
to  determine completely the basic photoproduction amplitudes and
the relationships between spin observables
 are addressed.  Such theorems are  most readily
demonstrated by representing all observables as bilinear products of helicity
amplitudes and using known properties of Dirac gamma
and spin-1 matrices.  The general angular dependence of spin
observables,
especially near thresholds and resonances, is examined for the
vector meson case.
  The criteria for a complete set
of observables  and the relationships between
observables are then presented.
\end{abstract}

\narrowtext

\section{Introduction}
\label{sec:introduction}

The photoproduction and electroproduction of mesons is  of
renewed interest now that
CEBAF has arrived.  The thresholds for $\rho,
 \omega$ and
$\phi$ mesons
 will be readily attained and exceeded.  The new high-intensity
continuous beams, and the development of polarized targets and
beams,
 will allow measurement of spin
observables with unprecedented precision.
  In addition, the measurement of
recoil baryons, including the $\Lambda$
 (which due to its weak decay $\Lambda \rightarrow p+\pi^-$
is spin self-analyzing),
 will provide even more spin observables.
Vector meson decays
 $(\rho \rightarrow   \pi~\pi ;\
 \omega \rightarrow \pi~\pi~\pi;\
 \phi \rightarrow K^+ K^-)$
provide a  measure of their density matrices,
 which is equivalent to determining the
intensity, polarization and tensor polarization of
 these vector mesons.  Such
measurements were made in the 1960's~\cite{oldexp}. With CEBAF,
 data of higher precision and completeness should be achievable~\cite{newexp}.

It is therefore timely and important to study the
 photo and electro-production of vector mesons.
 The associated form factors, and intermediate isobar states should
 test quark models.
The $\phi$ meson is of particular interest because of its
 simple ($\overline{s} s$) $^3S_{1^-}$ quark content.
  It is the positronium of strange quarks and, in parallel to
charmonium
($\overline{c}c$) states, should provide insights into
 basic QCD dynamics in the
nonperturbative region.  Perhaps an intimate connection between
production of
$\bar{s}s$  pairs and baryon polarization
 could provide detailed tests of quark descriptions.
 For example, the family of reactions
 $\gamma p \rightarrow K^+ \Lambda ,
 \gamma p \rightarrow K^+ \Sigma^0 ,
 \overline{p} p \rightarrow \phi \phi ,$
 and $\gamma p \rightarrow \phi p$
all involve $\bar{s}s$ strangeness production.
  In addition, they could also involve preexisting
$\bar{s}s$  content of the initial protons and antiprotons.
 That possibility has been
explored in a series of papers dedicated to finding
 direct evidence for  an
admixture of $\bar{s}s$  pairs in
baryons~\cite{henley1,henley2,snyang}.
 Spin observable are probably most sensitive to such
a configuration mixture.

In view of the above motivation, in this paper we discuss some
general
 features of spin observables
for the photoproduction of vector mesons.
  Our approach is similar to that used in
studies of pseudoscalar meson production and of the
 $\bar{p} p \rightarrow \bar{\Lambda} \Lambda$
 reaction near threshold~\cite{tabby1,FTS,ST}.
 However, for a vector meson the
spin-1 complicates the dynamics and a more general approach is
 required to keep track of
all  spin-observables and to demonstrate their general features, e.g.,
 their nodal structure and their normal energy evolution.
  For example,
the question of which spin observables vanish
 at $0^{\circ}$ and $180^{\circ}$, which have
nodes, where these nodes occur and how  they are expected
 to evolve with energy (based
on simple centrifugal barrier and/or resonance constraints)
are addressed here.
In addition, we examine the question of what constitutes a
 complete set of measurements,
e.g.,  which measurements are needed to determine the magnitude
 and phases of the basic
amplitudes.  Also, we wish to know which measurements are
 redundant, based on assumed
symmetries.  Such questions have been answered for the
 pseudoscalar meson case~\cite{walker,BDS1,BDS2}; those
 discussions are extended here to the vector meson case.
 Indeed,   generalized
discussions already exist  in the literature~\cite{simon,conzett},
which are however quite different from our treatment.

Our description uses a space of bilinear
 products of helicity amplitudes,
along with well-known properties of $4 \times 4$ and $3 \times 3$
matrix
 bases, to reveal the general features of spin observables.
We also use the idea of transversity amplitudes~\cite{morav}.
  For clarity, we also include the pseudoscalar case using this bilinear
form description and show how it  generalizes to the vector
 meson case and indeed to many other reactions.

\section{The Basic Amplitude and Spin Observables}

For vector meson photoproduction,
 $\vec{\gamma} + \vec{N} \rightarrow \vec{V} +
\vec{N'}$, our basic amplitude is of the form
\begin{equation}
   {\cal F} \equiv \left<\vec{q}\ \lambda_V\ \lambda_2 \mid T \mid \vec{k}\
\lambda\
 \lambda_1 \right> ,
\end{equation}
where the variables and the coordinate system conventions are
presented
 in Fig.~\ref{gamrho}.  The incident and final relative momenta
are $\vec{k}$ and $\vec{q},$ respectively.
  Jacob \& Wick (JW)~\cite{JW} phase conventions are used
throughout this paper.
 We use $\lambda = \pm 1$ to describe the
two transverse helicity states of the initial photon,
 $\lambda_1 = \pm 1/2$ to describe the
target$(N)$ proton  helicity,
 $\lambda_V = \pm 1, 0$ for the final vector meson  helicity
and $\lambda_2$ for the final$(N')$ baryon  helicity.
 For a real photon, the above
amplitude represents $3 \times 2 \times 2 \times 2 = 24$ complex
numbers.
However, by virtue of parity invariance there are 12 relations
between
 these amplitudes
 and consequently
 we have twelve independent complex
helicity amplitudes or 24 real numbers, at each energy and angle.
  We denote these twelve helicity amplitudes by
\begin{equation}
\left< \vec{q}\ \lambda_V\ \lambda_2 \mid T \mid \vec{p}\
\lambda\
 \lambda_1 \right> \rightarrow H_{a \lambda_V} (\theta) ,
\end{equation}
where $a = 1 \cdots 4$ and $\lambda_V = \pm 1, 0,$ and the particular
 matrix element assignments are given below.
  The pseudoscalar case is recovered
by simply taking $\lambda_V \rightarrow 0$ and then we
 have $1 \times 2 \times 2 \times 2 = 8$ complex
numbers and, after considering parity symmetries,
we obtain the usual four independent  $(a = 1 \cdots 4)$
helicity amplitudes $H_a(\theta)$~\cite{walker}.

The vector meson amplitude can be displayed as a $6 \times 4$
matrix
 in helicity space
\begin{eqnarray}
 {\cal F} =&
\left(\begin{array}{ccccc}

  &  H_{2 1}&H_{1 1}\ \ \ \vline&\ \ H_{3 -1}&-H_{4 -1}  \\
  & H_{4 1}&H_{3 1}\ \ \ \vline&-H_{1 -1}&\ \ H_{2 -1}  \\ \hline
  & H_{2 0}&H_{1 0}\ \ \ \vline&-H_{3 0}&\ \ H_{4 0}  \\
  & H_{4 0}&H_{3 0}\ \ \ \vline&\ \ H_{1 0}&-H_{2 0}  \\ \hline
  & H_{2 -1}&H_{1 -1} \ \vline&\ \ H_{3 1}&-H_{4 1}  \\
  & H_{4 -1}&H_{3 -1} \ \vline&-H_{1 1}&\ \ H_{2 1}
\end{array} \right)  ,
\label{matrixh}
\end{eqnarray}where the JW parity relation:
\begin{eqnarray}
&& \left< \vec{q}\ \lambda_V \lambda_2 \mid T \mid \vec{p}\
\lambda\
\lambda_1 \right> = \nonumber \\
 (-1)^{\Lambda_f -\Lambda_i}&& \left< \vec{q}\ -\lambda_V
-\lambda_2 \mid T \mid \vec{p}\ -\lambda  -\lambda_1 \right>
\end{eqnarray}
has been used, where $\Lambda_i \equiv \lambda - \lambda_1$ and
 $\Lambda_f \equiv \lambda_V - \lambda_2$.
The rows of ${\cal F}$ are labeled by the final state
 helicities $( \lambda_V, \lambda_2 )$
 and the columns by the initial photon and proton helicities
 $( \lambda, \lambda_1  ).$~\footnote{ We often
use $\lambda_1 = \pm 1$ to designate
the nucleon's helicity,  which really
has the values $\lambda_1 = \pm 1/2 .$}  The helicity amplitudes in ${\cal F}$
are
defined by the following,
\begin{eqnarray}
  H_{1 \lambda_V} &\equiv&
  <\lambda_V , \lambda_2 =  +1/2   \mid
 T  \mid \lambda =1 , \lambda_{1}= -1/2     > ,\nonumber \\
 H_{2 \lambda_V} &\equiv&
  <\lambda_V , \lambda_{2}=  +1/2   \mid
 T \mid \lambda =1 , \lambda_{1}= +1/2     > ,\nonumber \\
H_{3 \lambda_V} &\equiv&
  <\lambda_V , \lambda_{2}=  -1/2   \mid
 T \mid \lambda =1 , \lambda_{1}= -1/2    > ,\nonumber \\
H_{4 \lambda_V} &\equiv&
  <\lambda_V , \lambda_{2}=  -1/2  \mid
 T \mid \lambda =1 , \lambda_{1}= +1/2    >  ,
\label{defineH} \end{eqnarray} along with the JW parity rules.
  The pseudoscalar meson case is described by the $\lambda_V = 0\
(2 \times 4)$
part of the above matrix.

 General spin observables, $ \Omega,$ for  vector meson
photoproduction
 can be expressed in the
following trace form:
\begin{equation}
\Omega = \frac{ {\rm Tr} \left[ {\cal F} (A_{\gamma} A_N) {\cal F}^{\dag}
(B_V B_{N'})
\right]}{{\rm Tr}\left[ {\cal F} {\cal F}^{\dag} \right]} ,
\label{traceform}
\end{equation}
where the trace is over spin-space helicity quantum numbers
$\lambda, \lambda_1, \lambda_V, \lambda_2.$
This basic expression for observables is an average
 over a  classical ensemble of particles in the beam.
Interference occurs only at the quantum level
for the basic two-body reaction.
  The matrices have the dimensions: ${\cal F}~(6 \times 4);\
A_{\gamma} A_{N}~(4 \times 4);\ {\cal F}^{\dag} (4 \times 6);$
\  and $B_V B_{N'}~(6 \times 6).$

Here $A_{\gamma}$ denotes the usual $2 \times 2$
Hermitian spin matrices $(1, \vec{\sigma}_{\gamma}),$
 which describe the photon's two spin degrees of freedom.
 The $2 \times 2$ matrix
$A_N$ is similarly the $(1, \vec{\sigma}_N) $ spin matrices,
 as is $B_{N'} (1,
\vec{\sigma}_{N'}$), which describes the recoil baryon's spin state.
 The vector meson matrix
$B_V$ is a $3\times 3$ matrix and a complete set of these
  is provided by the $3 \times 3$ unit operator $(\openone)$,
  the usual spin-1 operators $\vec{S},$   plus five
 independent rank-2 operators
  $\tau_{2 \pm 2}, \tau_{2 \pm 1}$, and $\tau_{2 0}.$
 These operators are
 given in terms of the spin-1 spherical tensor operators
$$S_{1 \pm 1} = \mp \frac{S_x \pm i S_y}{\sqrt{2}} = \
 \mp \frac{S_{\pm}}{\sqrt{2}}, \hspace{0.5in}
S_{1 0} = S_z ,$$ by the tensor operator coupling
$\tau_{2 \mu} = [ S_1 \times S_1 ]_{2 \mu}\ .$

Associated with the above ensemble average, Eq.~\ref{traceform},
 there are density matrices describing the
polarization state of each particle.
  For example, the incident photon and the target proton
are described by:
\begin{equation}
\rho_{\gamma} = \frac{I_\gamma}{2} (\openone_\gamma +
\vec{\sigma}_{\gamma} \cdot \vec{P}_S) ,
\label{densphoton}
\end{equation}and
\begin{equation}
\rho_N = \frac{I_N}{2} (\openone_N + \vec{\sigma}_N \cdot
\vec{P}_N) .
\label{densN}
\end{equation}
The $\vec{P}_S$ above is the Stokes vector,  which is used
to describe the photon's polarization  as discussed in
Ref.~\cite{FTS}. The vector $\vec{P}_N$ describes the target polarization;
 its component in the $\hat{y}$ direction is denoted by
$T\equiv\vec{P}_N \cdot \hat{y}.$
 The final state density matrices are for the final baryon:
\begin{equation}
\rho_{N'} = \frac{I_{N'}}{2} (\openone_{N'} + \vec{\sigma}_{N'} \cdot
\vec{P}_{N'}) ,
\label{densNf}
\end{equation} and for the final vector meson:
\begin{equation}
\rho_V = \frac{I_V}{3} (\openone_V +   \vec{\bf{\cal S}} \cdot \vec{P}_V +
  3\ \tau \cdot T_V)  .
\label{densrho}
\end{equation}
 Using the definition
$\vec{\bf{\cal S}} \equiv \sqrt{\frac{3}{2}}\vec {S},$  gives us a vector meson
polarization
\begin{equation}
\vec{P}_V =   \frac{ {\rm Tr} \left[ {\cal F}{\cal F}^{\dag}\vec{{\cal S}}
  \right]}{{\rm Tr}\left[ {\cal F} {\cal F}^{\dag} \right]},
\label{vectorpolar}\end{equation}  for which each component is normalized to be
$\leq 1,$
 since $\vec{{\cal S}}\cdot\vec{{\cal S}}=3.$

The vector meson has not only a vector polarization, but also
a tensor polarization, $T_V^{2 \mu}.$  Here
 $\tau \cdot T_V \equiv \sum_\mu (-1)^\mu \tau_{2 -\mu} T_V^{2 \mu},$  for the
tensor polarization in a spherical tensor basis.  Then one has
\begin{equation}
T_V^{2 \mu} =    \frac{ {\rm Tr} \left[ {\cal F}{\cal F}^{\dag}\tau_{2 \mu}
  \right]}{{\rm Tr}\left[ {\cal F} {\cal F}^{\dag} \right]}.
\label{tensorpolar}
\end{equation}

The initial and final state density matrices are simply products of the
above forms,
$$\rho_i \equiv \rho_\gamma\ \times\ \rho_N; \hspace{0.5in}
 \rho_f \equiv \rho_V\ \times\  \rho_{N'}\ ,$$
where the helicity matrix elements of the above operators
are, for example, $< \lambda_1' \mid \rho_N \mid \lambda_1 >,$ etc.
Here the quantities $\vec{P}_S$, $\vec{P}_N$, $\vec{P}_{N'}$,
 $\vec{P}_V$ and $T_V^{2 \mu}$
 are real functions of angle and energy which
describe  how  the density matrix
   is formed from the basic matrices (or operators)
$ \openone, \vec{\sigma}_{\gamma}, \vec{\sigma}_N,
 \vec{\sigma}_{N'}, \vec{S}_V,$ and $\tau_{2 \mu}, $
which act in their respective helicity space.  These
``functions"  are called  spin observables, with
 the most familiar being the polarization.

To deduce the trace form(Eq.~\ref{traceform})
 for the spin observables, one begins by expressing the
general cross-section
in terms of the above density matrices; namely,
\begin{equation}
\sigma(\theta) = \varrho_0\ \frac{{\rm Tr} \rho_f}{{\rm Tr} \rho_i}
 ;
 \hspace{0.1in} {\rm where} \hspace{0.1in}
\rho_f  =  {\cal F}\ \rho_i\ {\cal F}^\dagger ,
\label{gencross}
\end{equation} and $\varrho_0 = q/(2k).$  Then, for a given
experimental situation,
one inserts the appropriate density matrices and extracts the
associated spin observable in trace form, Eq.~\ref{traceform}.

If one has an unpolarized beam and an unpolarized target,  the
initial density matrix is simply $\rho_i = I_\gamma I_N/4$ and
then with ${\rm Tr}\ \rho_i = I_\gamma I_N$ and
 $\rho_f =\frac{I_\gamma I_N}{4} ({\cal F}\ {\cal F}^\dagger)$
 {\it the} differential cross section is
\begin{equation}
\sigma(\theta) = \varrho_0\ \frac{1}{4}\ {\rm Tr} [ {\cal F}\ {\cal F}^\dagger
]
\equiv\  \varrho_0\ {\cal I}(\theta).
\label{diffcross}
\end{equation} Here we define the cross-section function ${\cal I}(\theta).$

 Another important case is the single spin observable called the
target polarization $T.$   Now one
has a polarized target and the initial density matrix is
$\rho_i = \rho_\gamma \times \rho_N = (I_\gamma I_N/4)\openone_\gamma
 ( \openone_N + \vec{\sigma}_{N} \cdot \vec{P}_{N}).$  and
the final density matrix is
$\rho_f = \rho_V \times \rho_{N'} = (I_V I_{N'}/4) .$
 Inserting these density matrices into Eq.~\ref{gencross},
 one finds that the target
polarization is given by:
\begin{equation}
 \Omega^{\alpha = 10 ,\beta=1} \rightarrow  T  =
 \frac{{\rm Tr} \left[ {\cal F}\   \openone_\gamma\
  \vec{\sigma}_N \cdot \hat{y}\  \
{\cal F}^\dagger\ \right]  }
 { {\rm Tr} \left[ {\cal F} {\cal F}^\dagger\ \right] } \ ,
\label{targetP}
\end{equation}where the labels $\alpha, \beta$ anticipate  a
subsequent discussion.

 Equation~{\ref{traceform} can be
deduced
in this same manner for all possible  experimental spin  setups,
including
not only single polarized spin cases, but also situations when two,
three,
or even, for the vector meson case,  four spins are setup as initial
states
or measured in the final state.  In all cases, one simply counts and
measures
a general cross section.  Expressions for extracting the spin
observables
from such cross-sections are available in the
literature~\cite{FTS,conzett}
 or simply derived from
the above density matrix formulation.

\section{Observables in Bilinear Helicity Product Form}

\subsection{ The BHP form and basis matrices}

Let us now map the above trace (or ensemble
 average) form over to a {\it bilinear helicity product} BHP form.
To cast the spin observables into a BHP form,  one needs to
insert the helicity amplitude matrix, Eq.~\ref{matrixh}, in the trace form
expression, Eq.~\ref{traceform}, with various choices of the
$A, B$ spin operators.~\footnote{The tedious algebra was done using
MAPLE.}
 Each spin observable $\Omega$ can  be
 written in the following general BHP product form:
\begin{equation}
\check{  \Omega}^{\alpha \beta} = \Omega^{\alpha \beta}\ {\cal I}
(\theta)
 = \pm \frac{1}{2}
\left< H \mid \Gamma^{\alpha} \omega^{\beta} \mid H \right> ,
\label{BHP}
\end{equation}
\begin{equation}
= \pm \frac{1}{2} \sum_{a, b,\ \lambda_V, \lambda'_{V} }
H^*_{a \lambda_V}\
\Gamma^{\alpha}_{a b}
\ \omega^{\beta}_{\lambda_V \lambda'_{V} }\  H_{b \lambda'_{V } }\ .
\label{helform}
\end{equation} See Eq.~\ref{diffcross} for the definition of
${\cal I}(\theta).$    This bilinear helicity product form,  which
 is the key expression for our analysis,
 often proves to be much more convenient
than the original ${\it trace\ form}$ of
Eq.~\ref{traceform}.~\footnote{ Note that
the ket notation $\mid H>$ does not refer to
a state vector,  but is adopted here to represent the helicity
amplitudes
with the convention that
 $<a \lambda_V \mid H> \equiv H_{a \lambda_V}(\theta).$}
 We use the symbol $\check{  \Omega}$
to indicate the above {\it profile function} form,  which is not to be
confused with the unit vector designation as in $\hat{y}.$

For any particular spin observable,  the
matrices $\Gamma^\alpha$ are one of  sixteen
  Hermitian $ 4\times4$ matrices  and
$\omega^\beta$ is selected from
  nine Hermitian $3\times3$ matrices.
 The Hermiticity of $\Gamma^\alpha $ and
$\omega^\beta$ assures that Eq.~\ref{helform} yields real
observables.
The $\Gamma$ matrices are Hermitian versions of the familiar
sixteen Dirac matrices:
$$\Gamma^{1\cdots 16}\ =  {\bf 1}, \gamma^0,
i\,\vec{ \gamma}, i\, \sigma^{0 x} , i\, \sigma^{0 y} ,
i\, \sigma^{0 z} ,
\sigma^{x y} ,
\sigma^{x z} ,
\sigma^{y z} ,$$ \begin{equation}
i\,\gamma^5 \gamma^0 ,
\gamma^5 \vec{\gamma} ,
 \gamma^5 .
\label{gammas} \end{equation}  Note that here the four dimensional
space is not that of a relativistic 4-spinor; instead, it is the
$a=1\cdots 4$ part of the helicity amplitude $H_{a \lambda_V}$
``space."

The vector meson part of the helicity amplitude ``space,"
i.e.,  the $\lambda_V \lambda'_{V}$ part, is a three-dimensional
space with nine associated Hermitian matrices $\omega^\beta.$
These are constructed from  nine spin-1 matrices
 as: $$\omega^\beta =
 {\bf 1}, \sqrt{3 / 2}\ \vec{S}\ ,
 \sqrt{6}\ t_{x y}\ ,
 \sqrt{6}\ t_{x z}\ ,
 \sqrt{6}\ t_{y z}\ , $$ \begin{equation}
  \sqrt{3}/2 ~[\pm~(t_{x x}~-~t_{y y})~-~\sqrt{3}~t_{z z}]\ ,
\end{equation}
where in a Cartesian basis:
 $$ t_{i j}~\equiv~(~S_i~S_j~+~S_j~S_i~)/2~-~(2/3)~\delta_{ij}\ \ .$$
The  matrices $t_{ij}$ are Cartesian versions of the earlier
operators $\tau_{2 \mu};$  however, we use a different notation since
$t$ acts in the helicity amplitude space (BHP) form of Eq.~\ref{BHP},
while $\tau_{2 \mu}$ appears in the trace form, Eq.~\ref{traceform};
see Appendix A.

The  matrices $\Gamma, \omega$ are defined to provide a complete Hermitian
basis
with the
properties: Tr$[\Gamma^\alpha\ \Gamma^\beta] = 4\ \delta_{\alpha
\beta},$\
 and
 Tr$[\omega^\alpha\ \omega^\beta] = 3\ \delta_{\alpha \beta}. $
The last five $\omega^{5,6,7,8,9}$ matrices are the
rank-2 operators associated with the vector meson's tensor polarization.
The properties of the matrices $\Gamma, \omega$
   are summarized in  appendices.

 Each value of the 16 superscripts  $\alpha  $ and
of the 9 superscripts
 $\beta$ label a particular  choice of
spin operators  $A_{\gamma}, A_N, B_V$ and $B_{N'}.$
  For example, the choice
$A_{\gamma} = A_N = B_V = B_{N'} = \openone $ leads to
the following BHP form
\begin{eqnarray}
\check{   \Omega}^{\alpha=1,  \beta=1} = {\cal I}(\theta) &=&
 \frac{1}{2} \sum_{a=1}^{4}\sum_{ \lambda_V= 0,\pm1} |H_{a \lambda}|^2
\nonumber \\
&=& \frac{1}{2} \left< H \mid \openone_N\ \openone_V \mid H \right>,
\end{eqnarray} which yields the differential cross
section using Eq.~\ref{diffcross}.
The product $\check{  \Omega}^{\alpha \beta} \equiv
\Omega^{\alpha \beta}\
{\cal I},$ which is called a profile function, is convenient since it
is proportional to bilinear products of helicity amplitudes.

Another example is $A_{\gamma} = B_V = B_{N'} = \openone,$
 and $A_N = \sigma^N_y$;
then, we recover the target polarization profile function
\begin{equation}
\check{  \Omega}^{\alpha=10,  \beta=1} \rightarrow \check{  T} =
 T\  {\cal I}(\theta) =
 -\frac{1}{2}  \left< H \mid \Gamma^{10}\ \omega^1 \mid H \right> ,
\end{equation} which motivated the earlier labels in Eq.~\ref{targetP}.

As one ranges over all possible choices of the operators $A$ and $B,$
 many of the traces vanish due to parity.
  There are sixteen values of $\alpha$, and
nine values of $\beta.$  Therefore,  we have
$16 \times 9 = 144$ possible $\Gamma^\alpha\ \omega^\beta$
products, which correspond to 144 distinct nonzero
 spin observables for the
 $\vec{\gamma} + \vec{N} \rightarrow \vec{V} + \vec{N'}$
 reaction, at each energy
 and angle.  For the pion case, there are only  sixteen
$(\alpha= 1 \cdots 16)$  such observables.
 Since for the vector(pseudoscalar) production
 we only need to know 23(7) of these
144(16) to make a complete measurement,
 there is clearly a great redundancy in the full
list of spin observables.
  Note that so far the only symmetry invoked has
 been parity.~\footnote{The rule for parity, based on
JW conventions, can be written as
$ {\cal P}_\gamma {\cal P}_P {\cal F} = {\cal F}^\dagger {\cal P}_N $
for the pion case and
${\cal P}_\gamma {\cal P}_P {\cal F} = {\cal F}^\dagger {\cal P}_V
{\cal P}_N
$ for the vector meson case, where ${\cal P}_V$ is an
 operator defined by its action on the
spin-1 eigenstates by
${\cal P}_V \mid 1\ \lambda_V > = (-1)^{1 -\lambda_V} \mid 1 -\lambda_V >.  $ }

\subsection{Single spin observables in BHP form}

The single spin observables for vector meson production are
now given explicitly in terms of the above $\Gamma$ and $\omega$
matrices.  Single spin observables involve experiments where
only one of the particles is polarized.
  We list first the case of no particles being polarized:
\begin{equation}
{\cal I} = \!\! \ \frac{1}{2} <H |\ \fbox{$\Gamma^1$}\ \
\fbox{$\omega^1$} |H >,
\label{BHP1}
\end{equation} which defines {\it the} cross-section and corresponds to
Eq.~\ref{diffcross}.  For convenience,  we count this
case as a single spin observables.

Matrices that are diagonal in the transversity
amplitude basis\footnote{Transversity amplitudes
are similar to helicity amplitudes except that
the axis of spin quantization is the transverse rather than the
particle's momentum direction.
Hence, transversity amplitudes are defined using the
axis of quantization perpendicular to the scattering plane
($\hat{  y}$)
instead of using the particle momentum
directions ($\hat{  z}, \hat{  z}'$). }
 are enclosed by a box in the above equations.
The significance of these boxed expressions and general
properties of single spin observables will be discussed later.

The first single spin observable is the target polarization  profile
$\check{T} \equiv \vec{P}_N \cdot \hat{y}\ {\cal I}$:
\begin{equation}
\check{  T} = \!\! \ - \frac{1}{2} <H |\ \fbox{$\Gamma^{10}$}\ \
\fbox{$\omega^1$}\
|H > ;
\label{BHP2}
\end{equation}  the next observable is the recoil polarization
$\check{P}_{N'} \equiv \vec{P}_{N'} \cdot \hat{y}'\ {\cal I}$, e.g.,  the
polarization of the
final baryon:
\begin{equation}
\check{  P}_{N'} = \!\! \ \frac{1}{2} <H |\ \fbox{$\Gamma^{12}$}\
 \fbox{$\omega^1$}\ |H > .
\label{BHP3}
\end{equation}

 The next single spin
 observable is the polarized photon
asymmetry $\check{\Sigma}  \equiv \vec{\Sigma}  \cdot \hat{x}\ {\cal I}$:
\begin{equation} \check{  \Sigma} = \!\! \ \frac{1}{2} <H |\ \fbox{$\Gamma^4$}
 \ \fbox{$\omega^A$}\  |H >.
\label{BHP4}
\end{equation}

 Now we turn to the final vector meson,  which has a vector
polarization $\check{P}_{V} \equiv \vec{P}_{V} \cdot \hat{y}'\ {\cal I}$:
\begin{equation}
\check{  P}_{V} = \!\! \ \frac{1}{2}  \sqrt{ \frac{3}{2}} <H | \openone  S_y
|H >
 = \!\! \  \frac{1}{2} <H |\ \fbox{$\Gamma^1$}\ \ \fbox{$\omega^3$}\  |H >,
\label{BHP5}
\end{equation} and also tensor polarizations,  for which the transverse
$(\hat{y})$
component is
\begin{eqnarray}
\check{  T}_{y y} & =&  \frac{1}{2}  <H |  \openone\ t_{y y} |H > \nonumber \\
              &=& \frac{1}{2}  <H |  \openone \ (S^2_y - \frac{2}{3}) |H >
\nonumber \\
              &=& \frac{1}{2}
 <H |\ \fbox{$\Gamma^1$}\ \
\left[ \frac{3\  \fbox{$\omega^A$} - \fbox{$\omega^1$} }{6}
\right]\ |H > ,
\label{BHP6}
\end{eqnarray}with the spherical tensor components given by:
\begin{equation}
\check{  T}_{2 1} = \!\! \ \sqrt{ \frac{1}{6} }\
<H |\ \fbox{$\Gamma^1$}\  \omega^6 |H > ,
 \label{BHP7}
\end{equation}
\noindent and
\begin{equation}
\check{  T}_{2 2} = \!\! \   \frac{1}{\sqrt{3}} \frac{1}{4}
 <H |\ \fbox{$\Gamma^1$}\  \left[  \omega^8 - \omega^9 \right]|H >   ,
 \label{BHP8}
\end{equation}
\begin{eqnarray}
\check{  T}_{2 0} &=&  -  \frac{1}{\sqrt{6}}
  <H |\ \fbox{$\Gamma^1$}\  \left[ \omega^8 + \omega^9 \right] |H > ,
\nonumber \\
 &=&  \frac{3}{2}  <H | t_{z z}  |H >  ,   \nonumber  \\
 &=&  \  \frac{3}{2}  <H | S^2_z - \frac{2}{3}  |H > .
\label{BHP9}
\end{eqnarray} Only eight of the above single spin
observables are independent,
 since we have $\check{T}_{2 \mu} = (-1)^\mu \check{T}_{2 -\mu}.$

The following characteristic combinations of the
$(3\times3)$ $\omega^{1,8,9}$ matrices are used above
and later:
\begin{eqnarray}
\left(\begin{array}{c }
\omega^A  \\
\omega^B  \\
\omega^C
\end{array} \right)
\equiv
\frac{1}{ 3}\left(
\begin{array}{ccc }
 1&(1-\sqrt{3})&(1+\sqrt{3})  \\
 (1-\sqrt{3})&(1+\sqrt{3})&1  \\
(1+\sqrt{3})&1&(1-\sqrt{3})
\end{array} \right)
\left(\begin{array}{c }
\omega^1  \\
\omega^8  \\
\omega^9
\end{array} \right) ,\label{comboC}
\end{eqnarray} which is a unitary transformation
in the $\omega$ basis.

\subsection{Double spin observables in BHP form}

 The double spin observables are now presented using the notation:
$$ \check{  C}_{i j}^{\gamma N} \equiv
  \frac{1}{2}\,< H |\ {  {\cal C}_{i j}^{\gamma N} }\ | H >.$$ The $\check{
C}$
is used to designate the spin observable profile function, while
the symbol ${\cal C}$ is used to designate a matrix
in the helicity  $H_{a \lambda_V} $ space. In addition,
superscripts are used to stipulate the two polarized
 particles in the reaction.  Finally, a matrix display
will be used to present the Cartesian components $(i, j)$
of the various spin observables; for example,
instead of writing ${\cal C}_{i j}$ as
${\cal C}_{x x},
 {\cal C}_{x y},
 {\cal C}_{x z}, $ these are presented below as the top
row of a matrix.  These conventions allow us to
present the 99 double spin observables in a relatively
compact form.~\footnote{ Of the 99 possible
double spin observables for vector meson photoproduction,
51 are nonzero, all others vanish by virtue of parity considerations.
 The nonzero 51 observables
 are not in general independent of the other spin observables.}

 \subsubsection{ Beam-target   observables}
 Let us start with the beam-target$(\gamma-N)$ double
spin observables.  The Cartesian terms are:
\begin{eqnarray}
{\cal C}^{\gamma N} =&
\!\! \   \left(\begin{array}{ccc}
0 \ & - \fbox{$\Gamma^{12}$}\  \fbox{$\omega^A$}\  & 0 \\
 & & \\
\Gamma^{5}\ \fbox{$\omega^A$}\ & 0\ & \Gamma^{3}\
\fbox{$\omega^A$}\ \\
 & & \\
\Gamma^{11}\ \fbox{$\omega^1$}\ \ &  0\ & \Gamma^{9}\
\fbox{$\omega^1$}\  \
\end{array} \right)  ,
\label{comboA}
\end{eqnarray} Note that there are 9 entries in the above Cartesian
component $(x,y,z)$ display,  but four vanish; namely,
${\cal C}_{x x}= {\cal C}_{x z}= {\cal C}_{y y}= {\cal C}_{z y}= 0.$
 Also note that a set of five independent $\Gamma^\alpha$ matrices
appear above
for $\alpha= 3,5,9,11,12,$ along with   $\omega^1$  and
  the linear combination  $\omega^A$ matrix (Eq.~\ref{comboA}).  Here both
 particles are in the initial state and the Cartesian components
refer to the initial $(x,y,z)$ axes of Fig.~\ref{gamrho}.  Of the nine possible
beam-target spin observables,  only five are nonzero.  Near the
vector meson production threshold,  only one of the
above double spin observables is nonzero; namely,
${\cal C}^{\gamma N}_{z z} \equiv  \Gamma^{9}\ \omega^1 ,$  see
Ref.~\cite{cetin2}.

\subsubsection{ Beam-recoil observables}

For the beam-recoil$(\gamma-N')$ double spin observables, we have a similar
display,
$\check{  C}_{i j}^{\gamma N'} \equiv
  \frac{1}{2}\,< H |\ {  {\cal C}_{i j}^{\gamma N'} } | H >$
\begin{eqnarray}
{\cal C}^{\gamma N'} =&
\!\! \  \left(\begin{array}{ccc}
0 \ & \fbox{$\Gamma^{10}$}\  \fbox{$\omega^A$}\   & 0 \ \\
 & & \\
\Gamma^{14}\ \fbox{$\omega^A$}\ & 0\ & -\Gamma^{7}\
\fbox{$\omega^A$}\  \\
 & & \\
 -\Gamma^{16}\ \fbox{$\omega^1$}\ \ & 0\ & -\Gamma^{2}\
\fbox{$\omega^1$}\  \
\end{array} \right) ,
\label{BR}
\end{eqnarray} but a different set of $\Gamma$ matrices appear;
namely,
  $\alpha= 2,7,10,14,16.$  The $\omega^1$ and $\omega^A$
matrices
 appear again as in Eq.~\ref{comboA}.
Here    the Cartesian components of the initial photon refers
  to the   $(x,y,z)$ axes of Fig.~\ref{gamrho},  while the final
axes $(x',y',z')$ are used for the recoil baryon $N'.$
Therefore,  we can extract a particular double spin observable
from Eq.~\ref{BR} as:
$$\check{  C}_{x y'}^{\gamma N'} = \frac{1}{2} <H \mid \Gamma^{10}
\omega^{A}  \mid H>
\equiv \check{\Omega}^{ 10,A},$$
where $$\check{\Omega}^{ 10,A} =  \frac{1}{3}\left[ \check{\Omega}^{ 10,1}
+ (1 -\sqrt{3}) \check{\Omega}^{ 10,8}
+ (1 +\sqrt{3})\check{\Omega}^{ 10,9} \right].$$ Of the nine possible
beam-recoil spin observables,  only five are nonzero.
 Near the
vector meson production threshold,  only two of the
above double spin observables are nonzero; namely,
${\cal C}^{\gamma N'}_{z z'} \equiv  - \Gamma^{2}\ \omega^1 ,$
and ${\cal C}^{\gamma N'}_{z x'} \equiv  - \Gamma^{16}\ \omega^1 ,$  see
Ref.~\cite{cetin2}.

\subsubsection{ Target-recoil observables}

For the target-recoil$( N-N')$ double spin observables, we have
yet another set of four $\Gamma$ matrices appearing:
$ \check{  C}_{i j}^{N N'} \equiv
  \frac{1}{2}\ < H |\ {  {\cal C}_{i j}^{N N'} }\ | H >
$ with
\begin{eqnarray}
{\cal C}^{N N'} =&
\!\! \  \left(\begin{array}{ccc}
 - \Gamma^{6}\ &  0 \ &  - \Gamma^{13}\   \\
 & & \\
0 \ & - \fbox{$\Gamma^4$}\   &  0\   \\
 & & \\
 \Gamma^{8}\ &  0 \ &  \Gamma^{15}\   \\
\end{array} \right) \ \fbox{$\omega^1$}\
\label{TR}   .
\end{eqnarray}
This time we have $\Gamma^\alpha$'s
with $\alpha= 4,6,8,13,15.$
Here    the Cartesian components of the initial nucleon $N$ refers
  to the   $(x,y,z)$ axes of Fig.~\ref{gamrho},  while the final
axes $(x',y',z')$ are used for the recoil baryon $N'.$
A particular double spin observable
from Eq.~\ref{TR} is:
$$\check{  C}_{y y'}^{N N'} = -\frac{1}{2} <H \mid \Gamma^4 \omega^1
\mid H>
\equiv \check{\Omega}^{ 4,1}.$$  This is often called $D_{n n} ,$  where
the
subscript $n$ refers to the normal to the scattering plane,
which in our case is $\hat{y}= \hat{y}'  .$  Of the nine possible
target-recoil spin observables,  only five are nonzero.
 Near the
vector meson production threshold,  all five of the
above double spin observables are nonzero, see Ref.~\cite{cetin2}.

\subsubsection{ Beam-vector meson observables}

We now consider the first double spin observable
 which involves both a polarized photon and
a final polarized vector meson. We again display the Cartesian
components as a  $i,j$ array, with the spin profile function
 $\check{  C}_{i j}$
constructed from the helicity space matrix ${\cal C}_{i j}$ as
$\check{  C}_{i j}^{\gamma V} \equiv
  \frac{1}{2}\,< H |{  {\cal C}_{i j}^{\gamma V} } | H > . $
 However, since the final vector meson has not only
a vector polarization,  but also five possible
tensor polarization components the matrix now involves the
following $3 \times 8$ matrix.  The three rows
refer to the
three Cartesian components $i=1,2,3$ or $x,y,z$ of the photon's
Stokes vector,  while the first three columns $j=1,2,3$ refer to the
vector meson's three polarization components $\vec{P}_{V}.$
The last five columns $j=4,5,6,7,8 $ refer to the tensor polarization of the
vector
meson $ \sqrt{6}\,T^V_{xy}, \sqrt{6}\,T^V_{xz}, \sqrt{6}\,T^V_{yz},
\frac{\sqrt{3}}{2}\,(\pm (T^V_{xx} + T^V_{yy}) - \sqrt{3} T^V_{zz}$
\widetext The  Cartesian components for the $\gamma$
refer to the original $(x, y, z)$ axes  and for the final vector
particle refer to the
final $(x',y',z')$ axes of Fig.~\ref{gamrho}.  The result is:

\begin{eqnarray}
{\cal C}^{\gamma V} =&
\begin{array}{ccccc}
( &   \fbox{$\Gamma^4$}\ &\fbox{$\Gamma^4$}\ &\Gamma^{1}  & )
\\
& &  & &  \\
& &  & &
\end{array}
\!\! \  \left(\begin{array}{cccccccc}
\ 0   \ & \fbox{$\omega^3$}\ \   &0 \   &0 \  &\omega^6 \  & 0
\  &\omega^B \  &\omega^C  \  \\
\  -\omega^7 \  &0 \  &\omega^5  \ &\omega^4 \  &0  \ &\omega^2 &
0  \ &
0  \  \\
\   \omega^2 \  &0 \  &\omega^4  \ &\omega^5 \  &0  \ &\omega^7 &
0  \ &
0  \
\end{array} \right) .
\label{BV}
\end{eqnarray}   The spin observable in the
 photon density matrix of Eq.~\ref{densphoton}
 has the three Cartesian components;
whereas, the vector meson density matrix
 Eq.~\ref{densrho} has three vector plus five tensor polarization
spin observables. Again, that is the origin of the
$3\times8$ nature of the above double spin observable display.
\narrowtext

For convenience, we separate
 the helicity space $H_{a \lambda_V}$
into separate $<\lambda_V\mid~\omega \mid~\lambda'_{V}>$
  and   $  \Gamma_{a b} $  matrices.
To illustrate our notation,   the following
 double spin observable
 $$ \check{ C}_{x y'}^{\gamma V } = \frac{1}{2} < H \mid
 \Gamma^4 \omega^3 \mid H >
 = \check{\Omega}^{ 4,3} $$ can be extracted from Eq.~\ref{BV}.
Another example is $$ \check{ C}_{z 4}^{\gamma V } = \frac{1}{2} < H \mid
 \Gamma^1 \omega^5 \mid H >
 = \check{\Omega}^{ 1,5} $$

 Of the 24 possible
beam-vector meson spin observables,  only 12 are nonzero.
 Near the
vector meson production threshold,  seven  of the
above twelve double spin observables are nonzero;
namely,
 $\check{ C}_{z, x'}^{\gamma  V},
\check{ C}_{z, z'}^{\gamma V },
\check{ C}_{y, 4}^{\gamma V },
\check{ C}_{x, 5}^{\gamma V },
\check{ C}_{y, 6}^{\gamma V },
\check{ C}_{x, 7}^{\gamma V },
\check{ C}_{x, 8}^{\gamma V },$
 see Ref.~\cite{cetin2}.

\subsubsection{Recoil-vector meson observables}

 Similarly,   the
recoil-vector meson $(N'-V)$ case  involves a final
polarized baryon$(N')$ and the $3 + 5$ vector plus tensor
components of the final vector meson.  In this case
the  Cartesian components for  both the $N'$ and the final vector
particle refer to the
final $(x,'y',z')$ axes of Fig.~\ref{gamrho}.  Thus, a $3\times8$
display appears again,  where the double polarization
spin observable $\check{  C}$ is expressed
in terms of a helicity amplitude space matrix ${\cal C}_{i j}:$
$\check{  C}_{i j}^{N'  V} \equiv
  \frac{1}{2}\, < H |{  {\cal C}_{i j}^{N' V} } | H >
$ with
\begin{eqnarray}
{\cal C}^{N' V} =&
\!\! \
\begin{array}{ccccc}
-( &
 \Gamma^{16}\ & -\fbox{$\Gamma^{12}$}\ &\Gamma^{2}
 & ) \\
& &  & &  \\
& &  & &
\end{array}
\!\! \  \left(\begin{array}{cccccccc}
\   \omega^2 \  &0 \  &\omega^4  \ &\omega^5 \  &0  \ &\omega^7 &
0  \ &
0  \  \\
\ 0   \ &\fbox{$\omega^3$}\ &0 \   &0 \  &\omega^6 \  & 0
\  &\omega^8 \  &\omega^9  \  \\
\   \omega^2 \  &0 \  &\omega^4  \ &\omega^5 \  &0  \ &\omega^7 &
0  \ &
0  \
\end{array} \right) .
\label{RV}
\end{eqnarray}To illustrate our notation again,   the following
 double spin observable
$$\check{  C}_{x', z'}^{  N'  V}
  =  -\frac{1}{2}
 < H \mid \Gamma^{16} \omega^4 \mid H >
\equiv \check{\Omega}^{ 16,4} $$ can be extracted from Eq.~\ref{BV}.
 Another example is $$\check{  C}_{z', 4}^{  N'  V}
  =  -\frac{1}{2}
 < H \mid \Gamma^{2} \omega^5 \mid H >
\equiv \check{\Omega}^{ 2,5} $$
 Of the 24 possible
recoil-vector meson spin observables,  only 12 are nonzero.
Near the
vector meson production threshold, all  of the
above twelve double spin observables are nonzero,
 Ref.~\cite{cetin2}.

\subsubsection{ Target-vector meson observables}

 Finally,
 the target-vector meson $(N-V)$ case has the same type of display
 with the double spin observable  related to a helicity
amplitude space matrix by
$$ \check{  C}_{i j}^{N   V} \equiv
  \frac{1}{2}\,< H |{  {\cal C}_{i j}^{N  V} } | H >,
$$ with again a $3\times 8$ matrix
\begin{eqnarray}
{\cal C}^{N  V} =&
\!\! \
\begin{array}{ccccc}
-( &
  \Gamma^{11}\ & \fbox{$\Gamma^{10}$}\ & \Gamma^{9}
 & ) \\
& &  & &  \\
& &  & &
\end{array}
\!\! \  \left(\begin{array}{cccccccc}
\   \omega^2 \  &0 \  &\omega^4  \ &\omega^5 \  &0  \ &\omega^7 &
0  \ &
0  \  \\
\ 0   \ &\fbox{$\omega^3$}\  \   &0 \   &0 \  &\omega^6 \  & 0
\  &\omega^8 \  &\omega^9  \  \\
\   \omega^2 \  &0 \  &\omega^4  \ &\omega^5 \  &0  \ &\omega^7 &
0  \ &
0  \
\end{array} \right) .
\label{TV}
\end{eqnarray}

 Of the 24 possible
target-vector meson spin observables,  only 12 are nonzero.
 Near the
vector meson production threshold,  ten of the
above twelve double spin observables are nonzero;
namely,
 $
\check{  C}_{x x'}^{N V},
\check{  C}_{z x'}^{N  V},
\check{  C}_{y y'}^{N V},
\check{  C}_{x z'}^{N  V},
\check{  C}_{z z'}^{N  V},
C^{NV}_{y,zz},C^{NV}_{x,yz},C^{NV}_{y,xz},C^{NV}_{y,xx},
 C^{NV}_{x,xy}
$
 see Ref.~\cite{cetin2}.

The significance of the boxed matrices will become clear when
we discuss the transversity amplitudes.

\subsubsection{ Some general remarks}

There are 99 possible double spin observables, which
  reduce  to 51 after use of parity conservation.
   The total number of nonzero observables near threshold
is 37  out of  these 51, see Ref.~\cite{cetin2}.
There are 18 single spin observables (we count the cross-section
as a single spin observable), which reduce  to 8 using parity.
Of these 8, five are nonzero near or at threshold, which include the
cross-section plus 4 vector meson spin observables; namely,
$T_{xx},T_{yy},T_{zz},$ and $T_{xz}.$  Since there are 12 complex
amplitudes, one needs  24-1=23 independent
measurements to determine the photoproduction
amplitudes completely, with one overall arbitrary phase.
 At first glance, it would seem that
with 8 single and 51 double spin observables, it suffices to do just
selected single and double spin measurements to completely
determine the 12 amplitudes.  This proves {\it not} to be the case.

 The reason for
this conclusion is that many of these spin observables yield redundant
information.
In Section V,  it is shown that spin observables
of the same ``phase class" can yield redundant information and that
one needs to go beyond single plus double spin observables to
obtain a complete set of experiments.  The ``phase class" will be defined
later.

Triple and  also quadruple spin observables have been derived
and complete the full set of $16 \times 9 =144$  spin observables.
The results are given in Appendix F. Before use of parity
there are 243 triple spin observables; after use of parity
conservation there are 123.
 Before use of parity
there are 216 quadruple spin observables; after use of parity
conservation there are 108.  Many of these double, triple and
quadruple spin observables involve linear combinations
or the same BHP forms
$\Gamma^\alpha \times \omega^\beta $
  that appear in other   single and double spin
observables.  That reveals many  relations between spin observables
and is  an important advantage of the BHP display.

The basic question is: which of these $8+51+123+108=290$
observables can and need to be
  measured to determine the 12 complex amplitudes?  Clearly,
  23 measurements are needed at each energy and angle,  but they have to
be selected to yield independent
information.  This problem is dealt with in Section V.
 In addition,  one can ask which
observables are expected to have nodes not only
at the $0^{\circ}$ and $180^{\circ}$ endpoints, but also
in between? That and related questions are addressed next.

\section{Nodal Structure--Legendre Class}

Having expressed the spin observables for vector meson
photoproduction
in  BHP form using the basic matrices $\Gamma$ and $\omega,$ we
can  analyze
these matrices for insights as to the ``nodal structure" of
observables.  Note that the original $\Gamma$ and $\omega$
matrices can be organized into groups according to their common
``shape;"
let us call these groupings ``Legendre classes."  For example,
the following matrices are of diagonal$(D)$ shape:
 $\Gamma^{1}$ $\Gamma^{2}$ $\Gamma^{9}$ $\Gamma^{15};$
whereas, the following are
antidiagonal $(AD)$:
 $\Gamma^{3}$ $\Gamma^{4}$
 $\Gamma^{6}$ $\Gamma^{7}.$
 The remaining eight matrices are either
of one class (called PL for left parallelogram form):
 $\Gamma^{10}$ $\Gamma^{11}$
 $\Gamma^{13}$ $\Gamma^{14},$
or of another class (called PR for right parallelogram shape):
 $\Gamma^{5}$ $\Gamma^{8}$
 $\Gamma^{12}$ $\Gamma^{16}.$
Similarly for the $\omega$ space, one has  matrices  of diagonal$(D)$
shape:
 $\omega^{1}$ $\omega^{4},$
antidiagonal$(AD)$ shape:
 $\omega^{5}$ $\omega^{A},$
crossed shape $(X)$:
 $\omega^{8}$ $\omega^{9},$
 $\omega^{B}$ $\omega^{C},$
and diamond or polygon $(P)$ shape:
 $\omega^{2}$ $\omega^{3},$
 $\omega^{6}$ $\omega^{7}.$
Explicit $\Gamma$ and $\omega$ matrices are presented in
Appendix B,
where they are grouped together by their common, $(D,AD,PL,PR)$-for
$\Gamma$ or
 $(D,AD,X, P)$-for $\omega,$ shapes.

   We stress  classification of the matrices by their
shape because classes of
 observables involving   $\Gamma$ and $\omega$ matrices
of the same shape
have angular dependencies given by related mixtures of the same
Wigner $d-$functions.
To illustrate this remark, let us consider the general spin observable
  $\check{\Omega}^{\alpha, \beta},$  which is given in terms of the
helicity amplitudes $H_{a, \lambda_V}(\theta)$ by Eq.~\ref{helform}.
The helicity amplitudes have the following
partial wave expansion
\begin{equation}
H_{a \lambda_V}(\theta) = \sum_{J_1}
  (2 J_1+1) H_{a \lambda_V}^{J_1}\ d^{J_1}_{  \Lambda_{a f} ,\Lambda_{a i}
}(\theta),
\label{partial1}
\end{equation}where $\Lambda_{a f}$ and $\Lambda_{a i}$ take on the following
values:
$$\Lambda_{a f} = \lambda_V - \zeta_a \ {\rm and}\ \Lambda_{a i} = \xi_a,$$
where
$\zeta_1 = \zeta_2 = - \zeta_3 = -\zeta_4 = 1/2$ and
$\xi_2 = \xi_4 = 1/2,  \xi_1=\xi_3 =3/2,$ see
Eq.~\ref{defineH}.  Using the above expansion for
$H_{a \lambda_V}$ and $ H_{b \lambda'_{V}}$ in
Eqs.~\ref{BHP},~\ref{helform},  we have:
\begin{eqnarray}
\check{  \Omega}^{\alpha \beta}(\theta)
&=& \pm \frac{1}{2} \sum_{a, b,\ \lambda_V, \lambda'_{V } }\sum_{J_1, J_2}
 H^{J_1 *}_{a \lambda_V}\  H^{J_2 }_{b\lambda'_V}\
\Gamma^{\alpha}_{a b}
\ \omega^{\beta}_{\lambda_V \lambda'_{V} } \nonumber \\
&& d^{J_1}_{\lambda_V-\zeta_a,\xi_a   }(\theta)
d^{J_2}_{ \lambda'_V-\zeta_b,\xi_b   }(\theta).
\label{partial2}
\end{eqnarray}
 This bilinear form can be combined to extract the
dependence on a single Wigner$-d.$  One finds:

\begin{eqnarray}
\check{  \Omega}^{\alpha \beta}(\theta)
&=&
\sum_{  {\cal J}, \Lambda,  \Lambda' }\
 \sum_{a, b,\ \lambda_V, \lambda'_{V } } ( 2 {\cal J} +1) \nonumber \\
&    & d^{{\cal J}}_{\Lambda,  \Lambda' }(\theta) \times
 \Xi^{ {\cal J},  \alpha \beta}_{\Lambda,  \Lambda' }\left[   a b ;
\lambda_V, \lambda'_{V } \right]\ ,
\label{partial3}
\end{eqnarray} with
\begin{eqnarray}
\Xi^{ {\cal J}, \alpha \beta }_{\Lambda,  \Lambda' }\left[  a b ;
\lambda_V, \lambda'_{V } \right]\
&\equiv&  \nonumber \\
  \pm \frac{1}{2} \sum_{J_1, J_2} \Gamma^{\alpha}_{a b}
\ \omega^{\beta}_{\lambda_V \lambda'_{V} }  &  &
H^{J_1 *}_{a \lambda_V}\  H^{J_2 }_{b \lambda'_V}      \nonumber \\
  \left(
\begin{array}{lll}
 J_1 & J_2 & {\cal J} \\
\lambda_V-\zeta_a&\lambda'_V-\zeta_b&\Lambda
\end{array}
\right) &&
  \left(
\begin{array}{lll}
 J_1 & J_2 & {\cal J} \\
\xi_a&\xi_b&\Lambda'
\end{array}
\right) \ .
\label{partial4}
\end{eqnarray} Note that $\Lambda$ depends on the
helicity labels $ a,b$ and $\lambda _V, \lambda'_V,$
while $\Lambda' $ depends only on the
helicity labels $ a,b.$

  To understand the consequence of the above
result, select a particular choice of spin observable by
designating the associated values of $\alpha,$ and $ \beta;$
 for example, take $10,1$ for the target polarization, see Eq.~\ref{targetP}.
 Now consider
the full family of  $\Gamma \times \omega$ matrices with
the same matrix ``shape" for both  $\Gamma$ and $\omega.$
That
family is called a ``Legendre class."
For the target polarization case, the matrix products
$\Gamma^{10,11,13,14} \times \omega^{1,4}$ are all of the
same shape and thus the associated spin observables form
a ``Legendre class."  (The
members of this target polarization class will be discussed later.)

  Since all members of a
``Legendre class"   vanish  for the same $(a,b;\lambda_V \lambda'_{V})$
values,  they are all formed from the same set of Wigner $d$-functions,
 $d^{{\cal J}}_{\Lambda,  \Lambda' }(\theta)$ of
 Eqs.~\ref{partial3}-\ref{partial4}.
If, for example,  every member of that set of Wigner $d$-functions
vanishes at $0^{\circ}$ and $180^{\circ},$  then every member of that
``Legendre class" of spin observables will also
vanish  at $0^{\circ}$ and $180^{\circ}.$  Similarly,
if  every member of that set of Wigner $d$-functions
has a zero or a node at $90^{\circ},$   then every member of that
``Legendre class" of spin observables will also
have a zero or a node at $90^{\circ}.$  These observations follow from the
fact that these families of spin observables are all expressed
by various combinations (see $\Xi,$ above)
 of the same set of Wigner $d$-functions.
The mixture coefficients, $\Xi,$ do depend on the partial wave helicity
amplitudes,  which is how dynamics of the reaction
affects the detailed angular dependencies.  If one truncates the
partial wave expansion due to either threshold or resonance considerations,
then the blend of Wigner $d$'s is strongly restricted and one can
demonstrate explicit associated angle dependencies of the
spin observable profiles.    For example,  if only one $J_1, J_2$
set of partial wave helicity amplitudes are nonzero,  then
using the triangle rule ${\cal J} =
 J_1 + J_2  \cdots J_1 - J_2,$  only a limited number of
${\cal J}$ values appear,  which severely restricts the
nodal structure. That  allows one to test and extract
specific dynamical information from the nodal structure and energy
evolution of spin observables (see Ref.~\cite{ST}).

Thus the angular dependencies of spin
observables can be grouped into classes\cite{FTS,ST} with the
same potential nodal structure (hence the nomenclature Legendre
class).
 To  examine the general role of
resonances on  the nodal structure of spin observables,  it is
most convenient to introduce explicit orbital angular momentum
quantum numbers.  Therefore,  in a separate paper Ref.~\cite{cetin2}  the
electric and magnetic multipole amplitudes for vector meson photoproduction
are studied in detail.

Another way of examining the angular dependence of a Legendre class
of spin observables is to return to the partial wave expansion,
Eq.~\ref{partial1}.
For a given Legendre class,  only selected values of $a$ and $\lambda_V$
appear.
Thus, one can pick the corresponding helicity amplitudes from the
first column of Table~\ref{wigtable}, then proceed to the partial wave
helicity amplitudes and the associated
Wigner $d-$ functions in the second and third columns.  The fourth
column gives the  range of $J$ needed for that
amplitude to contribute;  while the
last two columns indicate the value of the Wigner $d-$function
at the $0^\circ$ and $180^\circ$ endpoints.

\begin{table}
\caption{ Partial Wave expansion of the helicity amplitudes. The
associated Wigner $d-$functions are shown along with an indication of
their values at the $0^\circ$ and $180^\circ$ endpoints.
Six of the helicity amplitudes $H_{a, \lambda_V}(\theta)$
vanish at both endpoints; namely,
$H_{1,1}, H_{1,0}, H_{2, -1}, H_{3,0}, H_{3, -1}, H_{4,1} .$
Three of the helicity amplitudes
vanish only at the $0^\circ$ endpoint; namely,
$H_{1,-1}, H_{2,0}, H_{4, -1} .$
Three of the helicity amplitudes
vanish only at the $180^\circ$ endpoint; namely,
$H_{2,1}, H_{3,1}, H_{4,0} .$
  }
\begin{tabular}{cccccc}

$H_{a, \lambda_V}(\theta)$ & $H^J_{a, \lambda_V}$ &
$d^J_{\Lambda_f,\Lambda_i}(\theta)$ & $J$  & $0^\circ$& $180^\circ$   \\
&&&&&  \\
\tableline
&&&&& \\
$H_{1, 1}(\theta)$ & $H^J_{1, 1}$ &
$d^J_{1/2,3/2}(\theta)$ & $J\ge 3/2$  & $0 $& $ 0$   \\
&&&&&  \\
$H_{2, 1}(\theta)$ & $H^J_{2, 1}$ &
$d^J_{1/2,1/2}(\theta)$ & $J\ge 1/2$  & $1 $& $ 0$   \\
&&&&&  \\
$H_{3, 1}(\theta)$ & $H^J_{3, 1}$ &
$d^J_{3/2,3/2}(\theta)$ & $J\ge 3/2$  & $1 $& $ 0$   \\
&&&&&  \\
$H_{4, 1}(\theta)$ & $H^J_{4, 1}$ &
$d^J_{3/2,1/2}(\theta)$ & $J\ge 3/2$  & $0 $& $ 0$   \\
&&&&&  \\
\tableline
&&&&& \\
$H_{1, 0}(\theta)$ & $H^J_{1, 0}$ &
$d^J_{-1/2,3/2}(\theta)$ & $J\ge 3/2$  & $0 $& $ 0$   \\
&&&&&  \\
$H_{2, 0}(\theta)$ & $H^J_{2, 0}$ &
$d^J_{-1/2,1/2}(\theta)$ & $J\ge 1/2$  & $0 $& $ 1$   \\
&&&&&  \\
$H_{3, 0}(\theta)$ & $H^J_{3, 0}$ &
$d^J_{1/2,3/2}(\theta)$ & $J\ge 3/2$  & $0 $& $ 0$   \\
&&&&&  \\
$H_{4, 0}(\theta)$ & $H^J_{4, 0}$ &
$d^J_{1/2,1/2}(\theta)$ & $J\ge 1/2$  & $1 $& $ 0$   \\
&&&&&  \\
\tableline
&&&&& \\
$H_{1, -1}(\theta)$ & $H^J_{1, -1}$ &
$d^J_{-3/2,3/2}(\theta)$ & $J\ge 3/2$  & $0 $& $ 1$   \\
&&&&&  \\
$H_{2, -1}(\theta)$ & $H^J_{2, -1}$ &
$d^J_{-3/2,1/2}(\theta)$ & $J\ge 3/2$  & $0 $& $ 0$   \\
&&&&&  \\
$H_{3, -1}(\theta)$ & $H^J_{3, -1}$ &
$d^J_{-1/2,3/2}(\theta)$ & $J\ge 3/2$  & $0 $& $ 0$   \\
&&&&&  \\
$H_{4, -1}(\theta)$ & $H^J_{4, -1}$ &
$d^J_{-1/2,1/2}(\theta)$ & $J\ge 1/2$  & $0 $& $ 1$   \\
&&&&&
\end{tabular}
\label{wigtable}
\end{table}

 Using this table,
one can deduce which observables are zero at the endpoints.
For example,  consider the Legendre class of
observables for which both $\Gamma^\alpha$ and $\omega^\beta$
are diagonal, e.g.,   the class: $\Gamma^{1,2,9,15}\times \omega^{1,4}.$
These observables depend on linear combinations
of $\sum_{a\lambda_V}  \pm |H_{a,\lambda_V}|^2.$  From Table~\ref{wigtable}
it follows that the associated observables do not necessarily
vanish at the end points. In this diagonal
 Legendre class the single and double observables are:
 $ {\cal I}, {\cal C}^{\gamma V}_{z z'},{\cal C}^{\gamma N'}_{z z'},
{\cal C}^{N' V}_{z' z'}, {\cal C}^{\gamma N}_{z z }, {\cal C}^{N V}_{z z'}$
and  ${\cal C}^{N N'}_{z z'}.$  This information is extracted
 from Tables~\ref{matrixtoobs1}--\ref{matrixtoobs2},
where the relation between the $\alpha$ and $\beta$ choices of the
matrices are related to  explicit spin observables.

Another example of how to use Tables~\ref{wigtable}--\ref{matrixtoobs2}
   to deduce the
endpoint behavior of a given Legendre class of observables
is seen by examining the Legendre class of
observables for which  both $\Gamma^\alpha$   and $\omega^\beta$
are antidiagonal, e.g.,   the class: $\Gamma^{3, 4, 6, 7}\times \omega^{5,A}.$
Only the helicity amplitude products with $a,b = (1,4), (2,3)$
and $\lambda'_V =  -\lambda_V$ appear for these observables.
Now using Table~\ref{wigtable} with those products, it follows that
this Legendre class involved bilinear helicity products that vanish
at both $0^\circ$ and $180^\circ.$  From
Tables~\ref{matrixtoobs1}--\ref{matrixtoobs2},
we learn that the single, double and triple
spin observables of this class are:
 $\Sigma,
{\cal C}^{\gamma V}_{y z'}, {\cal C}^{\gamma N'}_{y z'},
{\cal C}^{\gamma N }_{y z}, {\cal C}^{N N' V}_{x y' 4},
{\cal C}^{N N' V}_{x z z'}, {\cal C}^{N N' V}_{y x' 4},
{\cal C}^{\gamma N' V}_{x z' z'}$ and ${\cal C}^{\gamma N N'}_{x z z'}.$

The procedure consists of using Tables~\ref{wigtable}--\ref{matrixtoobs2}
 along with the Legendre class information to determine the endpoint
rules.  Alternately,  one can use the general results
Eqs.~\ref{partial3}--\ref{partial4}.

\section{ Transversity, Phase Class and Complete Experiments }

\subsection{ Transversity}

\subsubsection{Pseudoscalar meson transversity}

 Let us deal with the question of which experiments are needed
to determine the magnitude and phase of the 12
helicity amplitudes for vector meson production.  Clearly,
23 experiments are needed at each energy and angle.  It is well
known that
for pseudoscalar meson photoproduction, it is possible to determine
the magnitude of the four transversity
 amplitudes by measuring all four
single spin observables.
 The remaining  three phases (one overall phase
is arbitrary) can be determined by selecting three additional
double spin observables, following the BDS~\cite{BDS1,BDS2} rules.
The pseudoscalar
meson case can be recovered from Eqs.~\ref{BHP1}--\ref{BHP4},
 by setting $\omega^A
\rightarrow 1$ and omitting all observables which involve
a final meson spin of 1.  In that limit, the BDS rules can be
understood
 by performing a unitary transformation on the
$\Gamma$ matrices along with a unitary transformation on the
helicity
amplitudes to generate the transversity amplitudes $|\widetilde{H}>:$
\begin{equation}
 |\widetilde{H}_a> \equiv \sum_{b}\ {\bf U}^{(4)}_{a, b} | H_b>,
\label{transvG1} \end{equation}and   a new set of $\Gamma$ matrices
\begin{equation}
 \widetilde{\Gamma}^\alpha \equiv {\bf U}^{(4)}\
 \Gamma^\alpha\ {{\bf U}^\dagger}^{(4)}  .
\label{transvG2} \end{equation}

 The
sixteen spin observables are
invariant under such a unitary
transformation in ``helicity space:"
\newline $$
\check{\Omega}^\alpha \propto\  <\,H\ | \Gamma^\alpha\,|\,H >
\equiv < \widetilde{H}\ | \widetilde{\Gamma}^\alpha\,|\ \widetilde{H}>.$$  The
physically
meaningful unitary operator is the
transversity choice,
 \begin{eqnarray}
{\bf U}^{(4)} =&
\!\! \ \frac{1}{2}  \left(\begin{array}{cccc}
1 \ & -i\  &\ \  i&\ \  1\\
1 \ &\ \  i\  & -i &\ \  1\\
1 \ &\ \  i\  &\ \  i & -1\\
1 \ & -i\  & -i & -1\
\end{array} \right)  ,
\end{eqnarray}   which involves rotating the helicity
quantization axis $(\hat{  z}$ and $\hat{  z}')$ to the direction
normal to the
scattering plane   $\hat{  y}=\hat{  y}', $ see Fig.~1.
With the above $4 \times 4$ unitary transversity transformation,
the following matrices now are diagonal:
 $$ \widetilde{\Gamma}^{1} \ \ \widetilde{\Gamma}^{4} \ \
 \widetilde{\Gamma}^{10} \ \ \widetilde{\Gamma}^{12};$$
  whereas, the following are now antidiagonal:
 $$\widetilde{\Gamma}^{2}\ \  \widetilde{\Gamma}^{7} \ \
 \widetilde{\Gamma}^{14}\ \  \widetilde{\Gamma}^{16}.$$
 The remaining eight matrices are either
of one class (called PL for left parallelogram form):
 $$\widetilde{\Gamma}^{6}\ \  \widetilde{\Gamma}^{8} \ \
 \widetilde{\Gamma}^{13}\ \  \widetilde{\Gamma}^{15},$$
or of another class (called PR for right parallelogram form):
 $$\widetilde{\Gamma}^{3}\ \  \widetilde{\Gamma}^{5} \ \
 \widetilde{\Gamma}^{9}\ \  \widetilde{\Gamma}^{11}.$$

\subsubsection{Vector meson transversity}

The above procedure can now be extended to the vector meson case
by introducing an additional unitary transversity
 transformation in
the $3 \times 3$ space
 \begin{eqnarray}
{\bf U}^{(3)} =&
\!\! \  \frac{1}{2} \left(\begin{array}{ccc}
-1 \ & \sqrt{2} i\  & 1 \\
-\sqrt{2} i \ & 0\  & -\sqrt{2} i \\
1 \ & \sqrt{2} i\  & -1 \
\end{array} \right)  ,
\label{transvO} \end{eqnarray}  which makes the
$\hat{  y}$-axis the quantization axis for the spin-1 meson.
Correspondingly, there is now
a transformation in ``helicity space:" $ <H\ | \Gamma^\alpha\
\omega^\beta |\ H >
\equiv  < \widetilde{H}\ | \widetilde{\Gamma}^\alpha\
 \widetilde{\omega}^\beta |\ \widetilde{H}>,$
with $$ |\widetilde{H}_{a, \lambda_V} > \equiv
 \sum_{b, \lambda'_{V}}\ {\bf U}^{(4)}_{a, b}\
{\bf U}^{(3)}_{\lambda_V, \lambda'_{V} } | H_{b, \lambda'_{V} }>.$$
 For the vector meson part of the transversity $\widetilde{\omega}$ space,
 the following matrices now have diagonal $(D)$ form: \
 $\widetilde{\omega}^{1} \ \ \widetilde{\omega}^{3}$
 $\widetilde{\omega}^{A};$
\ antidiagonal $(AD)$ form: $\widetilde{\omega}^{6};$
\ crossed $(X)$ form:
 $\widetilde{\omega}^{8} \ \
 \widetilde{\omega}^{9}\ \
 \widetilde{\omega}^{B}\ \
  \widetilde{\omega}^{C};$
and diamond or polygon $(P)$ shaped form:\
 $\widetilde{\omega}^{2} \ \ \widetilde{\omega}^{4} \ \ \
  \widetilde{\omega}^{5} \ \ \widetilde{\omega}^{7}.$
We have  extended the definition of transversity amplitudes to
the case of a vector meson.  The original expressions for the
spin observable profiles $\check{\Omega}^{\alpha \beta}$
are of the same form as given in Eqs.~\ref{BHP1}--\ref{TV},
except that the helicity amplitudes $H_{a \lambda_V}$
 are replaced by the transversity amplitudes $\widetilde{H}_{a, \lambda_V}$,
and the matrices are replaced by
 $\Gamma^\alpha \rightarrow \widetilde{\Gamma}^\alpha$  and
$\omega^\beta \rightarrow \widetilde{\omega}^\beta .$  In this new
representation the diagonal terms are indicated by the
boxed matrices in  Eqs.~\ref{BHP1}--~\ref{TV}}.  The shapes of
the tranversity-transformed matrices are presented in  Appendices~C \& D.

\subsection{Phase class}

As discussed in Appendices  C \& D,  the shape of the $\widetilde{\Gamma},
\widetilde{\omega}$  allows us to group these matrices into
``phase classes."  As in the ``Legendre classes" of the original matrices
 $ \Gamma ,  \omega, $  the shapes are defined by where   nonzero
entries appear in the matrix.  The classification into:
diagonal, antidiagonal, left parallelogram, right parallelogram;
crossed, and polygon shapes  is of significance in that these shapes
select the contributing bilinear helicity products.  For example,
if in the product $\widetilde{\Gamma}^\alpha \widetilde{\omega}^\beta$
both matrices are diagonal,  then that observable depends
on linear combinations of the product:$ \mid\widetilde{H}_{a,\lambda_V}\mid^2.$
If on the other hand the matrix product
$\widetilde{\Gamma}^\alpha \widetilde{\omega}^\beta$ has an entry
at the location $a,b$ in the $\widetilde{\Gamma}$ space
and at the location $\lambda_V,\lambda'_V$ in the
 $\widetilde{\omega}$ space,  then that observable depends on
the following product:
$$  \widetilde{H}^*_{a, \lambda_V} \widetilde{H}_{b,\lambda'_V} =
\mid \widetilde{H}_{a, \lambda_V}\mid\ \mid \widetilde{H}_{b,\lambda'_V}\mid
 \exp i(\phi_{b, \lambda'_V} - \phi_{a, \lambda_V}), $$ where
$\widetilde{H}_{a,\lambda_V} =
\mid \widetilde{H}_{a, \lambda_V}\mid\ \exp i(\phi_{a, \lambda_V}),$ etc.
 Thus,  the
shapes of the matrices in the tranversity description tell us which
phases
$\phi_{b, \lambda'_V} - \phi_{a, \lambda_V}
 \equiv \phi^{b,a}_{ \lambda'_V \lambda_V}$
are needed to determine the associated spin observable.
    In Fig.~\ref{vectoramps},
the basic problem of determining the twelve amplitudes
is illustrated, where the lengths of the vectors correspond to
the magnitude of the transversity amplitudes,  and the phases
of these complex amplitudes are also shown.  To fix this diagram,
we need to determine 12 magnitudes and then 11 phases;  one overall
phase and the overall orientation of the
diagram of  Fig.~\ref{vectoramps} is arbitrary.  This situation is
a generalization of the pseudoscalar case,  which is described in
Appendix E and by Fig. 3.

\subsection{Complete experiments}

The phase classification of the transversity matrices are a guide to the task
of picking a complete set of experiments. Thus the procedure is
to first select experiments which
 give information about the magnitudes of the 12 transversity
amplitudes and then to pick experiments which yield nonredundant
phase information.  In the  pseudoscalar case, the single spin
observables yield the magnitudes of all four transversity amplitudes.
 Then three  double spin
observables, selected using the  BDS~\cite{BDS1,BDS2} rules,
 yield three phases.
The task is similar for the vector case,  but the result is more complicated.

To extract the magnitudes of the transversity amplitudes,  we
need to examine all observables for which both
$\widetilde{\Gamma}^\alpha$ and $ \widetilde{\omega}^\beta$
are of the diagonal phase class.  These observables are
produced by all products:
$\widetilde{\Gamma}^{1,4,10,12}\times \widetilde{\omega}^{1,3,A},$
see Appendix C and D.

In the vector meson case there are only eight independent
single spin observables,  so we learn that this case is not as favorable
as the pseudoscalar meson case,  where the single spin observables
sufficed to determine all four amplitude magnitudes.  Indeed,
the situation is that only six of the
vector meson photoproduction single spin observables are of
diagonal form; namely,
${\cal I}, \check{ T},\check{    P}_{N'}, \check{    \Sigma},\check{    P}_V$
and the tensor polarization $\check{    T}_{yy}.$  Thus we need to
turn to the double spin observables for the remaining six diagonal
phase class matrices reside. {\bf Therefore, for vector meson
photoproduction it is not
possible to determine the magnitudes of the twelve transversity
amplitudes by only measuring six single spin observables.}

\section{Conclusions}
Several conclusions can be drawn from
 describing spin observables for vector meson photoproduction in bilinear
helicity product form.
 Here one
essentially extends the BDS~\cite{BDS1,BDS2} rules to include the
vector meson degree of freedom.

The diagonal matrices in the transversity basis are indicated by
boxes in Eqs.~\ref{BHP1}--~\ref{TV}. For the corresponding observables,
the diagonal nature of the product
 $\widetilde{\Gamma}^\alpha \ \widetilde{\omega}^\beta$ means that those
observables depend on linear combinations of the squared
magnitude of the transversity amplitudes, e.g.,  they depend on
combinations of $\pm |\widetilde{H}_{i, \lambda_V}|^2 .$

Those experiments which depend only on the magnitude
of the transversity amplitude are the following six
single spin observables:
$${\cal I} , \check{  T}, \check{  P}_{N'}, \check{  \Sigma}, \check{
P_V},
\check{  T_{yy}}, $$  plus the following six
double spin observables:
$$\check{  C}^{\gamma N}_{x y}, \
 \check{  C}^{\gamma N'}_{x y}, \
 \check{  C}^{N N'}_{y y}, \
 \check{  C}^{\gamma V}_{x y}, \
 \check{  C}^{N V}_{y y}, \
 \check{  C}^{N' V}_{y y}. $$ Note the above $x$-component for the
photon beam corresponds to a photon linearly polarized
perpendicular to the scattering plane~\cite{FTS}.
In contrast to the pseudoscalar meson production case,
the magnitudes of the transversity amplitudes for the vector meson
case can not be determined by just single spin observable
measurements, one needs to also perform six double spin
measurements.
Only six of the eight independent single spin observables are of
diagonal form; the remaining two single spin observables provide
transversity amplitude
 phase information.
In addition, of those six double spin observables
three involve measuring the spin state of the final vector meson (via
its decay)
 $$\check{  C}^{\gamma V}_{x y}, \
 \check{  C}^{N V}_{y y}, \
 \check{  C}^{N' V}_{y y}, $$ two require
a polarized photon beam
$$ \
 \check{  C}^{\gamma N}_{x y}, \check{  C}^{\gamma N'}_{x y}  $$ and one
requires a polarized
target
and measurement of the spin state of the recoil final
baryon
$$ \check{  C}^{N N'}_{y y}. $$  That sets the task for determining the
magnitudes of the transversity amplitudes, which
provide the most convenient situation.  Note that near threshold,
the two spin observables $\check{  C}^{\gamma V}_{x y},$
    and $\check{  C}^{\gamma N}_{x y}$ vanish.

 Observables for which
both $\widetilde{\Gamma}^\alpha$ and $\widetilde{\omega}^\beta$
are not diagonal depend on transversity amplitude phase differences.
As an extension of the BDS rules for the pseudoscalar meson
case, these transversity phase-dependent observables form
phase classes in which some experiments provide redundant
phase information.  A graphical procedure for
analyzing the redundancy and phase class of experiments
for vector meson production is illustrated in Fig.~2.

The procedure consists of picking a set of phases between
transversity amplitudes
that correspond to a given phase class for the $\Gamma$ and $\omega$ matrices.
Then list all the $\Gamma \times \omega$ matrices of that same phase class.
 It is best to start with the two spin spin observables that are not
diagonal in the transversity basis; for example $T_{20},T_{21}.$
The corresponding experiments can be selected from Tables~II \& III,  where
only the number crresponding to the number of unknown phases need to be
performed.  In that process,  some experiments prove to be feasible,  some
are difficult if not impossible;  indeed, in some cases one needs to
go to triple spin observables.  In any event,  this selection procedure
can be used to answer the question of the  experiments needed for a full
determination of all phases and also which ones will be extremely difficult
to determine because of realistic experimental conditions.

In addition to these features of using
 the bilinear helicity product form, one can deduce many other
aspects of spin observables based on general knowledge
of the properties of the $\Gamma^\alpha\ \omega^\beta$ matrices.
For example, when a ``Lorentz" $(S\ \gamma^\mu S^{-1} =
\ell^\mu_\nu\ \gamma^\nu)$ transformation or a parity, chirality or
time reversal operation is performed on the
 $\Gamma^\alpha\  $and $ \omega^\beta$ matrices
in the ``helicity amplitude space."  they reveal linear relations
between spin observables.  If  Fierz transformations are made on the
 $\Gamma^\alpha\  $and $ \omega^\beta$ matrices, again in helicity
space,
then quadratic relationships between spin observables
 are readily deduced.

An important part of our analysis is the use of transversity
amplitudes.  Many other types of amplitudes can be defined.
For example, one can use the unitary transformations from the
Dirac $\gamma-$matrices to the chiral, Majorana or other
$\gamma-$matrices sets to define new amplitudes;  corresponding
unitary transformations in the vector meson helicity
 3-space can also be invoked.
Thus the amplitude basis is hardly unique and one can
 deduce other sets and therefore deduce other
spin observables as the ones to measure
to determine the magnitude and/or phases of these alternately
defined amplitudes.  However,  there is something very special
about the transversity amplitudes, which is closely related to
using the normal to the scattering plan as the spin-quantization axis.
The special property is that the transversity amplitudes make
the optimum number of single spin observables independent
of amplitude phases. Correspondingly,  phase information is relegated to
the more complicated  spin observables.  This is most dramatic for
the four single spin observables for pseudoscalar meson photoproduction
in that single spin observables
 provide the magnitude of all four transversity amplitudes.
  That simplification is also of benefit for vector mesons,
which suggests that there
are compelling reasons to
use transversity amplitudes.

We hope that the procedure described here will be useful in
ascertaining the information content and the nodal behavior of
vector meson spin observables and in planning experiments.
  Measurement of all spin observables needed to fully
determine the twelve amplitudes is probably not feasible,
although it is good to know what is needed for that
full task.  Even without a full
experimental determination of the photoproduction amplitudes,
it is possible to extract useful dynamical information,
which is not an uncommon situation in strong interaction
physics.

\acknowledgments

We wish to thank Drs. S.~N.~Yang, B.~Saghai and C.~Fasano for their
helpful comments.

\appendix
\section{Basis Matrices }
\subsection{Four $ \times$ four--matrices }

The basic properties of the 16 Hermitian $\Gamma$ matrices
are presented here. These are well-known matrices,  except
that we have made them Hermitian in order to generate real observables
when used in the 4-dimensional part of our helicity space amplitudes.
The basic properties are:
$$ {\rm Tr}[ \ \Gamma^\alpha\  \Gamma^\beta \ ] = 4\
\delta_{\alpha\ \beta}
  \ \
 \hspace{0.5in}
 \Gamma^{\alpha \dagger} = \Gamma^\alpha  $$ These permit one to
expand any $4\times4$ matrix $X$ in terms of the $\Gamma$'s
and to extract the expansion coefficients $C_\beta$ using:
$ X = \sum_{\beta} \ C_{\beta}\ \Gamma^{\beta}$ and
    $ C_{\beta}  =  \frac{1}{4}\ {\rm Tr}[ \ \Gamma^\beta\ X ]. $
The definition of the $\Gamma$ matrices in terms of
$\gamma^\mu, \gamma^5,  \sigma^{\mu \nu} \cdots$
is presented in Eq.~\ref{gammas}.

\subsection{ Three $ \times$ three--matrices }

The basic properties of the 9 Hermitian $\omega$ matrices
are presented here. These are  Hermitian in order to generate real observables
when used in the 3-dimensional part of our helicity space amplitudes.
The basic properties are:
$$ {\rm Tr }[\  \omega^\alpha\   \omega^\beta\  ] = 3\
\delta_{\alpha \beta}
  \ \ \hspace{0.8in} \omega^{\alpha \dagger} = \omega^\alpha  $$
As in the $4\times 4$ case,  these properties allow one to expand
a general $3\times 3$ matrix and to extract the associated
expansion coefficients.

The nine $\omega$ matrices are defined in the text as a unit matrix, plus three
vector spin-one matrices,  plus a rank-2 tensor. As given earlier
the Cartesian form of the rank-2, symmetric Hermitian tensor is:
$$t_{ij} = \frac{S_i S_j + S_j S_i}{2} - \frac{2}{3} \delta_{i j}.  $$
The associated spherical tensor form is:
$$ t_{20} = \sqrt{\frac{3}{2}}\  \tau_{zz} = - \frac{1}{\sqrt{6}} (
\omega^8 + \omega^9) ,$$
$$ t_{2 \pm 1 } = \mp (  t_{xz} \pm i t_{yz} )  = \mp
\frac{1}{\sqrt{6}} ( \omega^6 \pm i \omega^7), $$
$$ t_{2 \pm 2 } =  ( +\frac{t_{xx} - t_{yy}}{2}  \pm i t_{xy} ) =
\frac{1}{\sqrt{3}} ( \frac{\omega^8 -  \omega^9}{2}) \pm
\frac{i}{\sqrt{6}} \omega^5 ,
 $$
\noindent where these are also  expressed in terms of the
Cartesian rank-2 tensor or the $\omega^{5-9}$ matrices.

The following diagonal terms are of particular interest:
$$ t_{yy} = -\frac{t_{2 2} + t_{2 -2}}{2}
 - \frac{t_{2 0}}{\sqrt{6}}
= S^2_y - \frac{2}{3}, $$
$$ t_{xx} = +\frac{t_{2 2} + t_{2 -2}}{2}
 - \frac{t_{2 0}}{\sqrt{6}}
= S^2_x - \frac{2}{3}. $$  Here the
matrix $t_{i j}$ is mapped to the
rank 2 spherical tensor operator $t_{2 \mu};$
 it is used in the BHP form for spin observables.
The same rules apply to the operator $\tau_{2 \mu},$  which
appears in the trace form for spin observables.

\section{Original--Basis Matrices }

\subsection{Original four $ \times$ four  }
The sixteen $\Gamma$ matrices can be grouped into
four classes with four members in each class
according to their ``shape."  By shape,  we mean the
location of nonzero entries.  For the original(Dirac)
Hermitian
matrices the shapes are: diagonal$(D);$ antidiagonal$(AD);$
left parallelogram$(PL);$ and right parallelogram$(PR).$
 The first Legendre class
is of diagonal shape$(D)$ and has $ \Gamma^{ 1,2,9,15}$ as its members:
 {\small
$$\Gamma_{D}=\
  \left [\begin {array}{cccc} a&0&0&0\\
\noalign{\medskip}0&b&0&0
\\ \noalign{\medskip}0&0&c&0\\ \noalign{\medskip}0&0&0&d
\end {array} \right ] \ ;
  \begin {array}{ccccc}
 & a& b& c& d \\
 \Gamma_{1}\ &+1&+1&+1&+1 \\
 \Gamma_{2}\ &+1&+1&-1&-1 \\
 \Gamma_{9}\ &+1&-1&+1&-1 \\
 \Gamma_{15}& -1&+1&+1&-1
\end {array} \ .
$$}
The second Legendre class
is of antidiagonal shape$(AD)$ and has $ \Gamma^{3,4,6,7}$ as its members:
 {\small

$$\Gamma_{AD}=
  \left [\begin {array}{cccc} 0&0&0& a \\ \noalign{\medskip}0&0&
 b &0\\ \noalign{\medskip}0& c &0&0\\ \noalign{\medskip}
 d &0&0&0\end {array}\right ]\ ;
  \begin {array}{ccccc}
 & a& b& c& d \\
 \Gamma_{3}&+i&+i&-i&-i \\
 \Gamma_{4}&+1&-1&-1&+1 \\
 \Gamma_{6}&-1&-1&-1&-1 \\
 \Gamma_{7}&+i&-i&+i&-i
\end {array} \ .
$$}

The third Legendre class
is of left  parallelogram shape$(PL)$ and has
$ \Gamma^{ 10,11,13,14}$ as its members:
 {\small
$$\Gamma_{PL}=
  \left [\begin {array}{cccc}
                    0&a&0&0\\
 \noalign{\medskip} b&0&0&0\\
 \noalign{\medskip} 0&0&0&c \\
 \noalign{\medskip} 0&0&d&0
\end {array}\right ]\ ;
 \begin {array}{ccccc}
 & a& b& c& d \\
 \Gamma_{10}&+i&-i&+i&-i \\
 \Gamma_{11}&-1&-1&-1&-1 \\
 \Gamma_{13}&-1&-1&+1&+1 \\
 \Gamma_{14}& +i&-i&-i&+i
\end {array} \ .
$$}

The third Legendre class
is of right   parallelogram shape$(PR)$ and has
$ \Gamma^{ 10,11,13,14}$ as its members:
 {\small
$$ \Gamma_{PR}=
  \left [\begin {array}{cccc} 0&0& a &0\\
\noalign{\medskip}0&0&0&b \\
 \noalign{\medskip}c &0&0&0\\ \noalign{\medskip}0&
 d &0&0\end {array}\right ]\ ;
  \begin {array}{ccccc}
 & a& b& c& d \\
 \Gamma_{5}\ &+i&-i&-i&+i \\
 \Gamma_{8}\ &-1&+1&-1&+1 \\
 \Gamma_{12}&-i&-i&+i&+i \\
 \Gamma_{16}& +1&+1&+1&+1
\end {array} \ .
$$}  These shapes are important for the
determination of the angular dependence
of spin observables, e.g.,  their nodal structure.

 Later,  the associated shapes for the
transversity transformed matrices $\widetilde{\Gamma}$ will be
presented,  which are useful for the analysis of
a complete set of experiments.

\subsection{Original three $ \times$ three  }

The nine $\omega$ matrices can also be grouped into
four classes   according to their ``shape."  By shape,  we mean the
location of nonzero entries.  For the original
Hermitian
matrices, the shapes are: diagonal$(D);$ antidiagonal$(AD);$
diamond or polygon$(P );$ and crossed$(X).$
In this case there are two $D,$ one $AD,$
four $P,$ and two $X$ matrices,  which accounts for the nine $\omega^\beta$
matrices.  In addition, we classify the three matrices $\omega^{A,B,C},$
which are particular linear combinations of $\omega^{1,8,9},$ see
 Eqs.~\ref{comboA}--~\ref{comboC} .
 The matrix $\omega^{A } $ is antidiagonal; whereas
 $\omega^{ B } $ and $\omega^{C} $ are of crossed $X$ shape.
 The explicit forms are:
{\small
$$\omega_{D} =
\left [\begin {array}{ccc}
                  a&0&0\\
\noalign{\medskip}0&b&0 \\
\noalign{\medskip}0&0&c
\end {array}\right ] \ ;
\begin {array}{cccc}
 & a& b& c \\
\omega{1}  &+1&+1&+1 \\
\omega{4} &\frac{\sqrt {3}}{2} &0&-\frac{\sqrt {3}}{2}
\end {array}
$$
$$\omega_{AD} =
\left [\begin {array}{ccc}
                  0&0&a\\
\noalign{\medskip}0&b&0 \\
\noalign{\medskip}c&0&0
\end {array}\right ] \ ;
\begin {array}{cccc}
 & a& b& c \\
\omega{5}&-i\frac{\sqrt{6}}{2}&0&+i\frac{\sqrt{6}}{2} \\
\omega{A}&-1&+1&-1
\end {array} \ .
$$
$$\omega_{P} =
\left [\begin {array}{ccc}
                  0&a&0\\
\noalign{\medskip}b&0&c \\
\noalign{\medskip}0&d&0
\end {array}\right ] \ ;
\begin {array}{ccccc}
 & a& b& c & d \\
\omega{2}&+\frac{\sqrt{3}}{2}&+\frac{\sqrt{3}}{2}&+\frac{\sqrt{3}}{
2}
&+\frac{\sqrt{3}}{2} \\
\omega{3}&-i\frac{\sqrt{3}}{2}&+i\frac{\sqrt{3}}{2}&-
i\frac{\sqrt{3}}{2}
&+i\frac{\sqrt{3}}{2} \\
\omega{6}&+\frac{\sqrt{3}}{2}&+\frac{\sqrt{3}}{2}&-
\frac{\sqrt{3}}{2}
&-\frac{\sqrt{3}}{2} \\
\omega{7}&-
i\frac{\sqrt{3}}{2}&+i\frac{\sqrt{3}}{2}&+i\frac{\sqrt{3}}{2}
&-i\frac{\sqrt{3}}{2}
\end {array} \ .
$$
$$\omega_{X} =
\left [\begin {array}{ccc}
                  a&0&b\\
\noalign{\medskip}0&c&0 \\
\noalign{\medskip}d&0&e
\end {array}\right ] \ ;
\begin {array}{cccccc}
 & a& b& c & d & e \\
\omega{8}&-\frac{1}{2}&+\frac{\sqrt{3}}{2}&+1
&+\frac{\sqrt{3}}{2}& -\frac{1}{2} \\
\omega{9}&-\frac{1}{2}&-\frac{\sqrt{3}}{2}&+1
&-\frac{\sqrt{3}}{2}& -\frac{1}{2} \\
\omega{B}&-\frac{\sqrt{3}}{2}&+\frac{1}{2}&+1
&\frac{1}{2} &-\frac{\sqrt{3}}{2} \\
\omega{C}&+\frac{\sqrt{3}}{2}&+\frac{1}{2}&+1
&\frac{1}{2} &+\frac{\sqrt{3}}{2} \\
\end {array} \ .
$$ }The classification of these matrices are helpful in the
analysis of the angular dependence of spin observables,
e.g.,  of the ``Legendre class" and the associated nodal
structure.  The associated shapes for the
transversity transformed matrices $\widetilde{\omega}$,
  which are useful for
the analysis of a complete set of experiments,  will be presented later.

\section{Transversity--Basis Matrices }

Introducing the transversity amplitudes involves
a unitary transformation of the basis matrices,
 see Eqs.~\ref{transvG1}-~\ref{transvO}.
These transformed matrices are presented here for the
$4\times4,$ $\widetilde{\Gamma}$ and $3\times3,$ $\widetilde{\omega}$ cases.

\subsection{Transversity four $\times$ four}

After the transversity transformation,   the
sixteen $\widetilde{\Gamma}$ matrices still form four classes,
with four members in each class.  Since these matrices
are part of the analysis of which experiments are needed to
determine the magnitude and phase of the transversity amplitudes,
we refer to these as the  ``phase  class."

 The first phase class
is of diagonal shape$(D)$ and has $\widetilde{\Gamma}^{ 1,4,10,12}$ as its
members:
{\small
$$\widetilde{\Gamma}_{D}=
  \left [\begin {array}{cccc} a&0&0&0\\
\noalign{\medskip}0&b&0&0
\\ \noalign{\medskip}0&0&c&0\\ \noalign{\medskip}0&0&0&d
\end {array} \right ] \ ;
  \begin {array}{ccccc}
 & a& b& c& d \\
 \widetilde{\Gamma}_{1}\ &+1&+1&+1&+1 \\
 \widetilde{\Gamma}_{4}\ &+1&+1&-1&-1 \\
 \widetilde{\Gamma}_{10}&-1&+1&+1&-1 \\
 \widetilde{\Gamma}_{12}&-1&+1&-1&+1
\end {array} \ .
$$}
 The second phase class
is of antidiagonal shape$A(D)$ and has $\widetilde{\Gamma}^{ 2,7,14,16}$
as its members:
{\small
$$\widetilde{\Gamma}_{AD}=
  \left [\begin {array}{cccc} 0&0&0& a \\ \noalign{\medskip}0&0&
 b &0\\ \noalign{\medskip}0& c &0&0\\ \noalign{\medskip}
 d &0&0&0\end {array}\right ]\ ;
  \begin {array}{ccccc}
 & a& b& c& d \\
 \widetilde{\Gamma}_{2}\ &+1&+1&+1&+1 \\
 \widetilde{\Gamma}_{7}\ &-i&-i&+i&+i \\
 \widetilde{\Gamma}_{14}&-1&+1&+1&-1 \\
 \widetilde{\Gamma}_{16}&+i&-i&+i&-i
\end {array} \ .
$$} The third phase class
is of left parallelogram shape$(PL)$ and has
 $\widetilde{\Gamma}^{ 6,8,13,15}$ as its members:
{\small
$$\widetilde{\Gamma}_{PL}=
  \left [\begin {array}{cccc}
                    0&a&0&0\\
 \noalign{\medskip} b&0&0&0\\
 \noalign{\medskip} 0&0&0&c \\
 \noalign{\medskip} 0&0&d&0
\end {array}\right ]\ ;
 \begin {array}{ccccc}
 & a& b& c& d \\
 \widetilde{\Gamma}_{6}\ &-1&-1&+1&+1 \\
 \widetilde{\Gamma}_{8}\ &-i&+i&-i&+i \\
 \widetilde{\Gamma}_{13}&+i&-i&-i&+i \\
 \widetilde{\Gamma}_{15}& -1&-1&-1&-1
\end {array} \ .
$$}
 The fourth phase class
is of right parallelogram shape$(PR)$ and has
 $\widetilde{\Gamma}^{3,5,9,11}$ as its members:
{\small
$$ \widetilde{\Gamma}_{PR}=
  \left [\begin {array}{cccc} 0&0& a &0\\
\noalign{\medskip}0&0&0&b \\
 \noalign{\medskip}c &0&0&0\\ \noalign{\medskip}0&
 d &0&0\end {array}\right ]\ ;
  \begin {array}{ccccc}
 & a& b& c& d \\
 \widetilde{\Gamma}_{3}\ &-i&-i&+i&+i \\
 \widetilde{\Gamma}_{5}\ &+1&-1&+1&-1 \\
 \widetilde{\Gamma}_{9}\ &+1&+1&+1&+1 \\
 \widetilde{\Gamma}_{11}&+i&-i&-i&+i
\end {array} \ .
$$}

\section{Transversity Three $ \times$ Three  }

After the transversity transformation,   the
nine $\widetilde{\omega}$ matrices still form four classes.
 For these transversity
Hermitian
matrices, the shapes are still: diagonal$(D);$ antidiagonal$(AD);$
diamond or polygon$(P );$ and crossed$(X).$
In this case there are two $D,$ one $AD,$
four $P,$ and two $X$ matrices,
which accounts for the nine $\widetilde{\omega}^\beta$
matrices.  In addition, we classify the three matrices
$\widetilde{\omega}^{A,B,C},$
which are particular linear combinations of $\widetilde{\omega}^{1,8,9},$ see
 Eqs.~\ref{comboA}--~\ref{comboC}.
 The matrix $\widetilde{\omega}^{A } $ is now diagonal; whereas
 $\widetilde{\omega}^{B } $ and $\widetilde{\omega}^{C } $ are of crossed $X$
shape.

 The first phase class
is of diagonal shape$(D)$ and has
 $\widetilde{\omega}^{1,3,A}$ as its members:
{\small
$$\widetilde{\omega}_{D} =
\left [\begin {array}{ccc}
                  a&0&0\\
\noalign{\medskip}0&b&0 \\
\noalign{\medskip}0&0&c \ .
\end {array}\right ] \ ;
\begin {array}{cccc}
 & a& b& c \\
\widetilde{\omega}^1&+1&+1&+1 \\
\widetilde{\omega}^{3}&+\frac{\sqrt {6}}{2} &0&-\frac{\sqrt {6}}{2} \\
\widetilde{\omega}^{A}&+1&-1&+1
\end {array} \ .
$$}
The second phase class
is of antidiagonal shape$(AD)$ and has
 $\widetilde{\omega}^{6}$ as its sole member:
{\small
$$\widetilde{\omega}_{AD} =
\left [\begin {array}{ccc}
                  0&0&a\\
\noalign{\medskip}0&b&0 \\
\noalign{\medskip}c&0&0
\end {array}\right ] \ ;
\begin {array}{cccc}
 & a& b& c \\
\widetilde{\omega}{6}&+i\frac{\sqrt{6}}{2}&0&-i\frac{\sqrt{6}}{2}
\end {array} \ .
$$}
The third phase class
is of polygon shape$(P)$ and has
 $\widetilde{\omega}^{2,4,5,7}$ as its four members:
{\small
$$\widetilde{\omega}_{P} =
\left [\begin {array}{ccc}
                  0&a&0\\
\noalign{\medskip}b&0&c \\
\noalign{\medskip}0&d&0
\end {array}\right ] \ ;
\begin {array}{ccccc}
 & a& b& c & d \\
\widetilde{\omega}{2}&-\frac{\sqrt{3}}{2}&-\frac{\sqrt{3}}{2}&-
\frac{\sqrt{3}}{2}
&-\frac{\sqrt{3}}{2} \\
\widetilde{\omega}{4}&-i\frac{\sqrt{3}}{2}&+i\frac{\sqrt{3}}{2}&-
i\frac{\sqrt{3}}{2}
&+i\frac{\sqrt{3}}{2}  \\
\widetilde{\omega}{5}&-\frac{\sqrt{3}}{2}&-
\frac{\sqrt{3}}{2}&+\frac{\sqrt{3}}{2}
&+\frac{\sqrt{3}}{2} \\
\widetilde{\omega}{7}&-
i\frac{\sqrt{3}}{2}&+i\frac{\sqrt{3}}{2}&+i\frac{\sqrt{3}}{2}
&-i\frac{\sqrt{3}}{2} \ .
\end {array} \ .
$$}
The fourth phase class
is of crossed shape$(X)$ and has
 $\widetilde{\omega}^{8,9,B,C}$ as its four members:
{\small
$$\widetilde{\omega}_{X} =
\left [\begin {array}{ccc}
                  a&0&b\\
\noalign{\medskip}0&c&0 \\
\noalign{\medskip}d&0&e
\end {array}\right ] \ ;
\begin {array}{cccccc}
 & a& b& c & d & e \\
\widetilde{\omega}{8}&\frac{1 -\sqrt{3}}{4}&\frac{3 +\sqrt{3}}{4}&
-\frac{1 -\sqrt{3}}{2}&\frac{3 +\sqrt{3}}{4}&\frac{1 -\sqrt{3}}{4} \\
\widetilde{\omega}{9}&\frac{1 +\sqrt{3}}{4}&\frac{3 -\sqrt{3}}{4}&
-\frac{1 +\sqrt{3}}{2}&\frac{3 -\sqrt{3}}{4}&\frac{1 +\sqrt{3}}{4} \\
\widetilde{\omega}{B}&\frac{1 -\sqrt{3}}{4}&\frac{3 +\sqrt{3}}{4}&
+\frac{1 -\sqrt{3}}{2}&\frac{3 +\sqrt{3}}{4}&\frac{1 -\sqrt{3}}{4} \\
\widetilde{\omega}{C}&\frac{1 +\sqrt{3}}{4}&\frac{3 -\sqrt{3}}{4}&
+\frac{1 +\sqrt{3}}{2}&\frac{3 -\sqrt{3}}{4}&\frac{1 +\sqrt{3}}{4}
\end {array} \ .
$$}  These phase classes are useful in specifying the
experiments needed to determine the magnitude and phases
of the 12 complex transversity amplitudes.

\section{Pseudoscalar Mesons}
\normalsize
We can return to the case of pseudoscalar meson
photoproduction by omitting all vector meson spin observables
and by replacing all $\omega$ matrices by zero, except
for the $\omega^{1, A} \rightarrow 1$ case.  In addition,
the twelve amplitudes reduce to four: $H_{a, \lambda_V} \rightarrow H_{a} .$
This limit is equivalent to looking at the $\lambda_V  \rightarrow 0$
terms only.  For the pseudoscalar meson case, we  next present the
BHP  spin observable profiles.

\subsection{Single spin observables}
 There are four single spin observables for pseudoscalar meson
photoproduction, where we include the cross-section:
\begin{eqnarray}
{\bf CROSS-SECTION: }&I =& \!\! \ \frac{1}{2} <H |\fbox{$\Gamma^{1}$}\  |H > \\
{\bf TARGET: }& \check{T} =& \!\! \ - \frac{1}{2} <H |\fbox{$\Gamma^{10}$}\  |H
>   \\
 {\bf RECOIL: }& \check{P}_{N'} =& \!\! \ \frac{1}{2} <H |\fbox{$\Gamma^{12}$}\
 |H >  \\
 {\bf BEAM:  }& \check{\Sigma} =& \!\! \ \frac{1}{2} <H |\fbox{$\Gamma^{4}$}\
|H >
 \end{eqnarray}  Note that all of these single spin observables are diagonal
in the transversity amplitude case,  which is the  meaning of the
boxed $\Gamma$ matrices.  Therefore,  measurement of the four spin observables
yields the magnitudes of all four transversity amplitudes
 $\widetilde{H}_1 \cdots \widetilde{H}_4.$
  To determine the  amplitude phases,  one needs to measure double
spin observables.

\subsection{  Double spin observables }
There are four transversity amplitudes and hence four amplitude phases;
however, one overall phase is arbitrary.  Therefore,  one needs to
perform three measurements to fix these three phases, see
Fig.~\ref{vectorpions}.

\begin{eqnarray}
{\bf  BEAM - TARGET: }\ \ \check{  C}_{i j}^{\gamma N} =&
  \frac{1}{2}\ < H |{  {\cal C}_{i j}^{\gamma N} } | H >  \nonumber
\end{eqnarray}
\begin{eqnarray}
{\cal C}^{\gamma N} =&
\!\! \   \left(\begin{array}{ccc}
0 \ & - \fbox{$\Gamma^{12}$}\  & 0 \ \\
\Gamma^{5}\  \ & 0\ & \Gamma^{3}\  \ \\
\Gamma^{11}   \ &  0\ & \Gamma^{9}  \
\end{array} \right)
 = \!\! \  \left(\begin{array}{ccc}
0 \ & - P_{N'}  & 0 \ \\
H  \ & 0\ & G  \ \\
F   \ &  0\ & E  \
\end{array} \right) \nonumber
\end{eqnarray}

\begin{eqnarray}
{\bf  BEAM - RECOIL: }\ \ \check{  C}_{i j}^{\gamma N'} =&
  \frac{1}{2}\ < H |{  {\cal C}_{i j}^{\gamma N'} } | H >  \nonumber
\end{eqnarray}
\begin{eqnarray}
{\cal C}^{\gamma N'} =&
\!\! \  \left(\begin{array}{ccc}
0 \ & \fbox{$\Gamma^{10}$}\  & 0 \ \\
\Gamma^{14}\   \ & 0\ &  -\Gamma^{7}\  \ \\
 -\Gamma^{16}   \ & 0\ & -\Gamma^{2}  \
\end{array} \right)
 = \!\! \  \left(\begin{array}{ccc}
0 \ & -T  & 0 \ \\
O_{x'}  \ & 0\ & O_{z'}  \ \\
C_{x'}   \ &  0\ & C_{z'}  \
\end{array} \right)  \nonumber
\end{eqnarray}

\begin{eqnarray}
{\bf  TARGET - RECOIL: }\ \ \check{  C}_{i j}^{N N'} =&
  \frac{1}{2}\ < H |{  {\cal C}_{i j}^{N N'} } | H >  \nonumber
\end{eqnarray}
\begin{eqnarray}
{\cal C}^{N N'} =&
\!\! \  \left(\begin{array}{ccc}
 - \Gamma^{6}\ &  0 \ &  - \Gamma^{13}\   \\
0 \ & \fbox{$ - \Gamma^{4}$}\  & 0 \ \\
 \Gamma^{8}\ &  0 \ &  \Gamma^{15}\   \\
\end{array} \right)
 = \!\! \  \left(\begin{array}{ccc}
T_{x'} \ & 0  & T_{z'} \ \\
0  \ & -\Sigma & 0  \ \\
L_{x'}   \ &  0\ & L_{z'}  \
\end{array} \right)  \nonumber
\end{eqnarray}

The three boxed $\Gamma$ matrices in the above double
spin observables, already appeared in the single spin observables.
Thus there are only four,  instead of five, members in each of the
above double spin categories.  In particular,  the following
double spin observables are equal to single spin observables:
$${\cal C}^{\gamma N}_{x y}= - P_{N'}\ \ {\cal C}^{\gamma N'}_{x y'}= -T \ \
  {\cal C}^{N N'}_{y y'}=  -\Sigma .$$ These equalities are direct
consequences of parity conservation.  For the vector meson case,
the appearance of the $\omega^A$ matrix in the corresponding
double spin observables yields linear relations between observables,
 not the above restriction;  hence,  in that case there
are five double spin observables in these categories.

 In addition, the $\Gamma^\alpha$
matrices for each double spin observable are of the same ``phase
class." For ${\cal C}^{\gamma N}$ $\alpha= 3,5,9,11$ appear, which are
of phase class $PR.$  Based on this shape
category,these observables depend on the following
relative phases: $\phi_{1,3},\phi_{2,4}.$
For ${\cal C}^{\gamma N'}$ $\alpha= 2,7,14,16$ appear, which are
of phase class $AD.$  Based on this shape
category,these observables depend on the following
relative phases: $\phi_{1, 4},\phi_{2,3}.$
For ${\cal C}^{N N'}$ $\alpha= 6,8,13.15$ appear, which are
of phase class $PL.$ Based on this shape
category,these observables depend on the following
relative phases: $\phi_{1,2},\phi_{3,4}.$
Here,  we label the phases using $\phi_{a,b}$ for the phase
difference $\phi_{a,b}=\phi_{a }-\phi_{ b}$ between the
transversity amplitudes $\widetilde{H}_a$ and   $\widetilde{H}_b,$  see
Fig.\ref{vectorpions}.

Making three measurements of the same phase class observables would
be redundant;  only two are needed for the two phases.  The third measurement
should be taken from another phase class.  Thus one needs three
double spin measurements,  but not more than two from a given phase class.
 Here we have presented a derivation of the
BDS~\cite{BDS1,BDS2} rules,  based on the shape of the
 $\widetilde{\Gamma}$ matrices
and the geometric picture of the transversity amplitudes shown in
 Fig.\ref{vectorpions}.  The advantage of this rendition of the BDS
theorem is that it can be generalized to the case of vector meson
photoproduction,  see Fig.\ref{vectoramps}.

\subsection{ Triple spin observables}

In the pseudoscalar meson case,  the following triple
spin observables can be derived in the BHP form.  Again,  the
triple spin observables are displayed in a Cartesian
format with:
\begin{eqnarray}
\check{  C}_{ x i j}^{\gamma N N'} =&
  \frac{1}{2}\ < H |{  {\cal C}_{i j}^{\gamma N N'} } | H >  ,
\end{eqnarray}
\begin{eqnarray}
\check{  C}_{ y i j}^{\gamma N N'} =&
  \frac{1}{2}\ < H |{  {\cal C}_{ y i j}^{\gamma N N'} } | H >  ,
\end{eqnarray}
\begin{eqnarray}
\check{  C}_{ z i j}^{\gamma N N'} =&
  \frac{1}{2}\ < H |{  {\cal C}_{ z i j}^{\gamma N N'} } | H >  ,
\end{eqnarray}

\begin{eqnarray}
{\cal C}^{\gamma N N' }_{x i j} =&
\!\! \
\!\! \  \left( \begin{array}{ccc}
  \Gamma^{15}\ & 0\ &  -\Gamma^{8}   \\
0 \ & \fbox{$ -\Gamma^{1}$}\  & 0 \ \\
  \Gamma^{13}\ & 0\ & - \Gamma^{6}
\end{array} \right)   \
 = \!\! \  \left(\begin{array}{ccc}
-L_{z'} \ & 0  & L_{x'} \ \\
0  \ & -I & 0  \ \\
T_{z'}   \ &  0\ & -T_{x'}  \
\end{array} \right)   , \nonumber
\end{eqnarray}

\begin{eqnarray}
{\cal C}^{\gamma N N' }_{y i j} =&
\!\! \
\!\! \  \left( \begin{array}{ccc}
  0\ & - \Gamma^{9}\ & 0\   \\
 - \Gamma^{2}\ & 0\ &  \Gamma^{16}   \\
  0\ & -\Gamma^{11}\ & 0\
\end{array} \right)   \
 = \!\! \  \left(\begin{array}{ccc}
0 \ & -E  & 0 \ \\
C_{z'}  \ & 0 & -C_{x'}  \ \\
0   \ &  F\ & 0  \
\end{array} \right)   , \nonumber
\end{eqnarray}

\begin{eqnarray}
{\cal C}^{\gamma N N' }_{z i j} =&
\!\! \
\!\! \  \left( \begin{array}{ccc}
  0\ & - \Gamma^{3}\ & 0\   \\
   \Gamma^{7}\ & 0\ &  \Gamma^{14}   \\
  0\ &  \Gamma^{5}\ & 0\
\end{array} \right)   \
 = \!\! \  \left(\begin{array}{ccc}
0 \ & G  & 0 \ \\
-O_{z'}  \ & 0 & O_{x'}  \ \\
0   \ &  H\ & 0  \
\end{array} \right)  .  \nonumber
\end{eqnarray}  Here all three particles with spin are involved.
Of the $3\times9$ possible triple spin observables, 15
are nonzero.  Three of these are equal to spin spin observables
and the remaining 12 are equal to double spin
observables.  Therefore,  there is no new information
in triple spin observables for pseudoscalar meson photoproduction,
and it is fortunately not necessary to consider such complicated
measurements.  This is not the case for vector
meson photoproduction.

\section {Vector Mesons  }

The single and double spin observables were presented in the text.
Using the BHP approach and MAPLE,  it is possible to derive
explicit expressions for the triple and quadruple
spin observables for  vector
meson photoproduction.
 Triple spin observables involve the spin
of three particles,  including the vector and tensor polarization
of the vector meson.

\subsection{Triple spin observables }

 There are four types of triple spin observables.
The first three types,   $N, N',V,$
  $\gamma, N',V,$
and   $\gamma, N,V$ involve the vector meson.
The fourth type involves the $\gamma,N, N' $ particles and
does not include the vector meson:
\begin{eqnarray}
{\cal C}^{N N' V}_{x i j} =&
\!\! \
\begin{array}{ccccc}
( &
  \Gamma^{6}\ & -\Gamma^{3}\ & \Gamma^{13}
 & ) \\
& &  & &  \\
& &  & &
\end{array}
\!\! \  \left(\begin{array}{cccccccc}
\ 0   \ &\fbox{$\omega^3$}\  \   &0 \   &0 \  &\omega^6 \  & 0
\  &\omega^8 \  &\omega^9  \  \\
\   \omega^2 \  &0 \  &\omega^4  \ &\omega^5 \  &0  \ &\omega^7 &
0  \ &
0  \  \\
\ 0   \ &\fbox{$\omega^3$}\  \   &0 \   &0 \  &\omega^6 \  & 0
\  &\omega^8 \  &\omega^9  \
\end{array} \right)    , \  \nonumber
\end{eqnarray}

\begin{eqnarray}
{\cal C}^{N N' V}_{y i j} =&
\!\! \
\begin{array}{ccccc}
( &
  \Gamma^7\ & -\fbox{$\Gamma^4$}\ &\Gamma^{14}
 & ) \\
& &  & &  \\
& &  & &
\end{array}
\!\! \  \left(\begin{array}{cccccccc}
\   \omega^2 \  &0 \  &\omega^4  \ &\omega^5 \  &0  \ &\omega^7 &
0  \ &
0  \  \\
\ 0   \ &\fbox{$\omega^3$}\  \   &0 \   &0 \  &\omega^6 \  & 0
\  &\omega^8 \  &\omega^9  \  \\
\   \omega^2 \  &0 \  &\omega^4  \ &\omega^5 \  &0  \ &\omega^7 &
0  \ &
0  \
\end{array} \right)    ,   \  \nonumber
\end{eqnarray}

\begin{eqnarray}
{\cal C}^{N N' V}_{z i j} =&
\!\! \
\begin{array}{ccccc}
-( &
  \Gamma^8\ & -\Gamma^5\ &\Gamma^{15}
 & ) \\
& &  & &  \\
& &  & &
\end{array}
\!\! \  \left(\begin{array}{cccccccc}
\ 0   \ &\fbox{$\omega^3$}\  \   &0 \   &0 \  &\omega^6 \  & 0
\  &\omega^8 \  &\omega^9  \  \\
\   \omega^2 \  &0 \  &\omega^4  \ &\omega^5 \  &0  \ &\omega^7 &
0  \ &
0  \  \\
\ 0   \ &\fbox{$\omega^3$}\  \   &0 \   &0 \  &\omega^6 \  & 0
\  &\omega^8 \  &\omega^9  \
\end{array} \right)  .   \
\end{eqnarray}

\begin{eqnarray}
{\cal C}^{\gamma N' V}_{x i j} =&
\!\! \
\begin{array}{ccccc}
( &
  \Gamma^{14}\ & \fbox{$\Gamma^{10}$}\ & -\Gamma^{7}
 & ) \\
& &  & &  \\
& &  & &
\end{array}
\!\! \  \left(\begin{array}{cccccccc}
\   \omega^7 \  &0 \  &-\omega^5  \ &\omega^4 \  &0  \ &-\omega^2
& 0  \ &
0  \  \\
\ 0   \ &\fbox{$\omega^3$}\  \   &0 \   &0 \  &\omega^6 \  & 0
\  &\omega^B \  &\omega^C  \  \\
\   \omega^7 \  &0 \  &-\omega^5  \ &\omega^4 \  &0  \ &-\omega^2
& 0  \ &
0  \
\end{array} \right)    ,   \   \nonumber
\end{eqnarray}

\begin{eqnarray}
{\cal C}^{\gamma N' V}_{y i j} =&
\!\! \
\begin{array}{ccccc}
( &
  \Gamma^{14}\ & -\fbox{$\Gamma^{10}$}\ & -\Gamma^{7}
 & ) \\
& &  & &  \\
& &  & &
\end{array}
\!\! \  \left(\begin{array}{cccccccc}
\ 0   \ &\fbox{$\omega^3$}\  \   &0 \   &0 \  &\omega^6 \  & 0
\  &\omega^B \  &\omega^C  \  \\
\   \omega^7 \  &0 \  &-\omega^5  \ &\omega^4 \  &0  \ &-\omega^2
& 0  \ &
0  \  \\
\ 0   \ &\fbox{$\omega^3$}\  \   &0 \   &0 \  &\omega^6 \  & 0
\  &\omega^B \  &\omega^C  \
\end{array} \right)    ,   \nonumber
\end{eqnarray}

\begin{eqnarray}
{\cal C}^{\gamma N' V}_{z i j} =&
\!\! \
\begin{array}{ccccc}
-( &
  \Gamma^{16}\ & -\fbox{$\Gamma^{12}$}\ & \Gamma^{2}
 & ) \\
& &  & &  \\
& &  & &
\end{array}
\!\! \  \left(\begin{array}{cccccccc}
\ 0   \ &\fbox{$\omega^3$}\  \   &0 \   &0 \  &\omega^6 \  & 0
\  &\omega^8 \  &\omega^9  \  \\
\   \omega^2 \  &0 \  &\omega^4  \ &\omega^5 \  &0  \ &\omega^7 &
0  \ &
0  \  \\
\ 0   \ &\fbox{$\omega^3$}\  \   &0 \   &0 \  &\omega^6 \  & 0
\  &\omega^8 \  &\omega^9  \
\end{array} \right)  .   \
\end{eqnarray}

\begin{eqnarray}
{\cal C}^{\gamma N V}_{x i j} =&
\!\! \
\begin{array}{ccccc}
-( &
  \Gamma^{5}\ & \fbox{$\Gamma^{12}$}\ & \Gamma^{3}
 & ) \\
& &  & &  \\
& &  & &
\end{array}
\!\! \  \left(\begin{array}{cccccccc}
\   \omega^7 \  &0 \  &-\omega^5  \ &\omega^4 \  &0  \ &-\omega^2
& 0  \ &
0  \  \\
\ 0   \ &\fbox{$\omega^3$}\  \   &0 \   &0 \  &\omega^6 \  & 0
\  &\omega^B \  &\omega^C  \  \\
\   \omega^7 \  &0 \  &-\omega^5  \ &\omega^4 \  &0  \ &-\omega^2
& 0  \ &
0  \
\end{array} \right)    ,   \    \nonumber
\end{eqnarray}

\begin{eqnarray}
{\cal C}^{\gamma N V}_{y i j} =&
\!\! \
\begin{array}{ccccc}
-( &
  \Gamma^{5}\ & -\fbox{$\Gamma^{12}$}\ & \Gamma^{3}
 & ) \\
& &  & &  \\
& &  & &
\end{array}
\!\! \  \left(\begin{array}{cccccccc}
\ 0   \ &\fbox{$\omega^3$}\  \   &0 \   &0 \  &\omega^6 \  & 0
\  &\omega^B \  &\omega^C  \  \\
\   \omega^7 \  &0 \  &-\omega^5  \ &\omega^4 \  &0  \ &-\omega^2
& 0  \ &
0  \  \\
\ 0   \ &\fbox{$\omega^3$}\  \   &0 \   &0 \  &\omega^6 \  & 0
\  &\omega^B \  &\omega^C  \
\end{array} \right)    ,   \   \nonumber
\end{eqnarray}

\begin{eqnarray}
{\cal C}^{\gamma N V}_{z i j} =&
\!\! \
\begin{array}{ccccc}
-( &
  \Gamma^{11}\ & \fbox{$\Gamma^{10}$}\ & \Gamma^{9}
 & ) \\
& &  & &  \\
& &  & &
\end{array}
\!\! \  \left(\begin{array}{cccccccc}
\ 0   \ &\fbox{$\omega^3$}\  \   &0 \   &0 \  &\omega^6 \  & 0
\  &\omega^8 \  &\omega^9  \  \\
\   \omega^2 \  &0 \  &\omega^4  \ &\omega^5 \  &0  \ &\omega^7 &
0  \ &
0  \  \\
\ 0   \ &\fbox{$\omega^3$}\  \   &0 \   &0 \  &\omega^6 \  & 0
\  &\omega^8 \  &\omega^9  \
\end{array} \right) .   \
\end{eqnarray}

\begin{eqnarray}
{\cal C}^{\gamma N N' }_{x i j} =&
\!\! \
\!\! \  \left( \begin{array}{ccc}
  \Gamma^{15}\ & 0\ &  -\Gamma^{8}   \\
  0\ & - \Gamma^{1 }\ & 0\   \\
  \Gamma^{13}\ & 0\ & - \Gamma^{6}
\end{array} \right)   \  \fbox{$\omega^A$}    ,  \nonumber
\end{eqnarray}

\begin{eqnarray}
{\cal C}^{\gamma N N' }_{y i j} =&
\!\! \
\!\! \  \left( \begin{array}{ccc}
  0\ & - \Gamma^{9}\ & 0\   \\
 - \Gamma^{2}\ & 0\ &  \Gamma^{16}   \\
  0\ & -\Gamma^{11}\ & 0\
\end{array} \right)   \  \fbox{$\omega^A$}    ,   \nonumber
\end{eqnarray}

\begin{eqnarray}
{\cal C}^{\gamma N N' }_{z i j} =&
\!\! \
\!\! \  \left( \begin{array}{ccc}
  0\ & - \Gamma^{3}\ & 0\   \\
   \Gamma^{7}\ & 0\ &  \Gamma^{14}   \\
  0\ &  \Gamma^{5}\ & 0\
\end{array} \right) .   \ \fbox{$ \omega^1$}
\end{eqnarray}

\subsection{Quadruple spin observables}

  Quadruple spin observables involve the spin
of all four particles,  including the vector and tensor polarization
of the vector meson.  Of course all of these involve
the vector meson, including its vector and tensor polarization;
hence, the $3\times 8$  Cartesian display appears again.
The question arises: Are all of the quadruple spin-observables
redundant?  In the pseudoscalar meson case the full case
of triple spin observables were redundant in that they were all determined by
single and/or double spin observable measurements.  In the vector
meson case, we have:

\widetext
\begin{eqnarray}
{\cal C}^{ \gamma N N' V}_{x x i j} =&
\!\! \
\begin{array}{ccccc}
( &
  \Gamma^{15}\ & -\Gamma^{9}\ & -\Gamma^{8}
 & ) \\
& &  & &  \\
& &  & &
\end{array}
\!\! \  \left(\begin{array}{cccccccc}
\ 0   \ &\fbox{$\omega^3$}\  \   &0 \   &0 \  &\omega^6 \  & 0
\  &\omega^B \  &\omega^C  \  \\
\   \omega^7 \  &0 \  & -\omega^5  \ &\omega^4 \  &0  \ &
-\omega^2 & 0  \ &
0  \  \\
\ 0   \ &\fbox{$\omega^3$}\  \   &0 \   &0 \  &\omega^6 \  & 0
\  &\omega^B \  &\omega^C  \
\end{array} \right)   , \  \nonumber
\end{eqnarray}

\begin{eqnarray}
{\cal C}^{ \gamma N N' V}_{x y i j} =&
\!\! \
\begin{array}{ccccc}
- ( &
  \Gamma^{2}\ & \fbox{$\Gamma^{1}$}\ & \Gamma^{16}
 & ) \\
& &  & &  \\
& &  & &
\end{array}
\!\! \  \left(\begin{array}{cccccccc}
\   \omega^7 \  &0 \  & -\omega^5  \ &\omega^4 \  &0  \ &
-\omega^2 & 0  \ &
0  \  \\
\ 0   \ &\fbox{$\omega^3$}\  \   &0 \   &0 \  &\omega^6 \  & 0
\  &\omega^B \  &\omega^C  \  \\
\   \omega^7 \  &0 \  & -\omega^5  \ &\omega^4 \  &0  \ &
-\omega^2 & 0  \ &
0  \
\end{array} \right)   , \  \nonumber
\end{eqnarray}

\begin{eqnarray}
{\cal C}^{ \gamma N N' V}_{x z i j} =&
\!\! \
\begin{array}{ccccc}
( &
  \Gamma^{13}\ & -\Gamma^{11}\ & -\Gamma^{6}
 & ) \\
& &  & &  \\
& &  & &
\end{array}
\!\! \  \left(\begin{array}{cccccccc}
\ 0   \ &\fbox{$\omega^3$}\  \   &0 \   &0 \  &\omega^6 \  & 0
\  &\omega^B \  &\omega^C  \  \\
\   \omega^7 \  &0 \  & -\omega^5  \ &\omega^4 \  &0  \ &
-\omega^2 & 0  \ &
0  \  \\
\ 0   \ &\fbox{$\omega^3$}\  \   &0 \   &0 \  &\omega^6 \  & 0
\  &\omega^B \  &\omega^C  \
\end{array} \right)   , \  \nonumber
\end{eqnarray}

\begin{eqnarray}
{\cal C}^{ \gamma N N' V}_{y x i j} =&
\!\! \
\begin{array}{ccccc}
( &
   -\Gamma^{15}\ & \Gamma^{9}\ & \Gamma^{8}
 & ) \\
& &  & &  \\
& &  & &
\end{array}
\!\! \  \left(\begin{array}{cccccccc}
\   \omega^7 \  &0 \  & -\omega^5  \ &\omega^4 \  &0  \ &
-\omega^2 & 0  \ &
0  \  \\
\ 0   \ &\fbox{$\omega^3$}\  \   &0 \   &0 \  &\omega^6 \  & 0
\  &\omega^B \  &\omega^C  \  \\
\   \omega^7 \  &0 \  & -\omega^5  \ &\omega^4 \  &0  \ &
-\omega^2 & 0  \ &
0  \
\end{array}  \right)  , \  \nonumber
\end{eqnarray}

\begin{eqnarray}
{\cal C}^{ \gamma N N' V}_{y y i j} =&
\!\! \
\begin{array}{ccccc}
 ( &
 - \Gamma^{2}\ &\fbox{$ \Gamma^{1}$}\ & \Gamma^{16}
 & ) \\
& &  & &  \\
& &  & &
\end{array}
\!\! \  \left(\begin{array}{cccccccc}
\ 0   \ &\fbox{$\omega^3$}\  \   &0 \   &0 \  &\omega^6 \  & 0
\  &\omega^B \  &\omega^C  \  \\
\   \omega^7 \  &0 \  & -\omega^5  \ &\omega^4 \  &0  \ &
-\omega^2 & 0  \ &
0  \  \\
\ 0   \ &\fbox{$\omega^3$}\  \   &0 \   &0 \  &\omega^6 \  & 0
\  &\omega^B \  &\omega^C  \
\end{array} \right)   , \  \nonumber
\end{eqnarray}

\begin{eqnarray}
{\cal C}^{ \gamma N N' V}_{y z i j} =&
\!\! \
\begin{array}{ccccc}
( &
  -\Gamma^{13}\ & -\Gamma^{11}\ & \Gamma^{6}
 & ) \\
& &  & &  \\
& &  & &
\end{array}
\!\! \  \left(\begin{array}{cccccccc}
\   \omega^7 \  &0 \  & -\omega^5  \ &\omega^4 \  &0  \ &
-\omega^2 & 0  \ &
0  \  \\
\ 0   \ &\fbox{$\omega^3$}\  \   &0 \   &0 \  &\omega^6 \  & 0
\  &\omega^B \  &\omega^C  \  \\
\   \omega^7 \  &0 \  & -\omega^5  \ &\omega^4 \  &0  \ &
-\omega^2 & 0  \ &
0  \
\end{array} \right)   , \  \nonumber
\end{eqnarray}

\begin{eqnarray}
{\cal C}^{ \gamma N N' V}_{z x i j} =&
\!\! \
\begin{array}{ccccc}
( &
  \Gamma^{6}\ & -\Gamma^{3}\ & \Gamma^{13}
 & ) \\
& &  & &  \\
& &  & &
\end{array}
\!\! \  \left(\begin{array}{cccccccc}
\   \omega^2 \  &0 \  & \omega^4  \ &\omega^5 \  &0  \ & \omega^7
& 0  \ &
0  \  \\
\ 0   \ &\fbox{$\omega^3$}\  \   &0 \   &0 \  &\omega^6 \  & 0
\  &\omega^8 \  &\omega^9  \  \\
\   \omega^2 \  &0 \  & \omega^4  \ &\omega^5 \  &0  \ & \omega^7
& 0  \ &
0  \
\end{array} \right)   , \  \nonumber
\end{eqnarray}

\begin{eqnarray}
{\cal C}^{ \gamma N N' V}_{z y i j} =&
\!\! \
\begin{array}{ccccc}
 ( &
  \Gamma^{7}\ & - \fbox{$\Gamma^{4}$}\ & \Gamma^{14}
 & ) \\
& &  & &  \\
& &  & &
\end{array}
\!\! \  \left(\begin{array}{cccccccc}
\ 0   \ &\fbox{$\omega^3$}\  \   &0 \   &0 \  &\omega^6 \  & 0
\  &\omega^8 \  &\omega^9  \  \\
\   \omega^2 \  &0 \  & \omega^4  \ &\omega^5 \  &0  \ & \omega^7
& 0  \ &
0  \  \\
\ 0   \ &\fbox{$\omega^3$}\  \   &0 \   &0 \  &\omega^6 \  & 0
\  &\omega^8 \  &\omega^9  \
\end{array} \right)   , \  \nonumber
\end{eqnarray}

\begin{eqnarray}
{\cal C}^{ \gamma N N' V}_{z z i j} =&
\!\! \
\begin{array}{ccccc}
( &
 - \Gamma^{8}\ & \Gamma^{5}\ & -\Gamma^{15}
 & ) \\
& &  & &  \\
& &  & &
\end{array}
\!\! \  \left(\begin{array}{cccccccc}
\   \omega^2 \  &0 \  & \omega^4  \ &\omega^5 \  &0  \ & \omega^7
& 0  \ &
0  \  \\
\ 0   \ &\fbox{$\omega^3$}\  \   &0 \   &0 \  &\omega^6 \  & 0
\  &\omega^8 \  &\omega^9  \  \\
\   \omega^2 \  &0 \  & \omega^4  \ &\omega^5 \  &0  \ & \omega^7
& 0  \ &
0  \
\end{array} \right)   . \
\end{eqnarray}

\clearpage
 \begin{figure}[t]
\caption{
The coordinate system and kinematical variables for vector meson
photoproduction. Here V denotes the vector meson
and $\lambda_V$ its helicity.
}
\label{gamrho}
\end{figure}
\begin{figure}
\caption{This diagram displays the
  magnitudes and selected relative phase angles of the
 twelve transversity amplitudes $\widetilde{H}_{1 \lambda_V}
 \cdots \widetilde{H}_{4 \lambda_V}$
 (with $\lambda_V =0, \pm1)$ for
photoproduction of vector mesons.
 The twelve complex amplitudes are determined within an overall phase
if the twelve magnitudes and eleven angles are fixed by 23
appropriate measurements. The magnitudes are determined by measuring
  six single spin observables plus six
double spin observables, see the text.  Then independent phase angle
measurements are made by selecting  spin observables from
different ``phase class" observables.  The phase angles
are labeled by the convention $\phi^{s t}_{\lambda_V \lambda'_V}$   }
\label{vectoramps}
\end{figure}

\begin{figure}
\caption{ This diagram displays the
  magnitudes and selected relative phase angles of the
 four transversity amplitudes $\widetilde{H}_1 \cdots \widetilde{H}_4$ for
photoproduction of pseudoscalar mesons.
 The four complex amplitudes are determined within an overall phase
if the four magnitudes and three angles are fixed by seven
appropriate measurements. The magnitudes are determined by measuring
the four single spin observables,  then independent phase angle
measurements are made by selecting three  double spin observables from
at least two different ``phase class" observables.
The phase angles
are labeled by the convention $\phi^{s t} $  }
\label{vectorpions}
\end{figure}
\newpage
\mediumtext

\begin{table}
\caption{ Spin Observables in BHP form as products of $\Gamma^\alpha,$ and
 $\omega^\beta $ matrices.  Here the $\alpha=1 \cdots 16,
  \beta= 1-9,A,B,C$ range is displayed.  Single, double and
triple, but not quadruple, spin observables are shown.  The entries
of ``phase class" $\widetilde{\Gamma}_{PL} \times \widetilde{\omega}_P$
do not appear--they are all quadruple spin observables. }
\begin{tabular}{ccccccc}

$\alpha\backslash\beta$&1&2&3&4&5&6  \\
&&&&&&  \\
\tableline
&&&&&&  \\
 1&
${\cal I}$&
${\cal C}^{\gamma V}_{zx'}$&
$P_V$&
${\cal C}^{\gamma V}_{zz'}$&
${\cal C}^{\gamma V}_{z4}$&
$\sqrt{\frac{2}{3}}T_{21}$\\
&&&&&&  \\
\tableline
&&&&&&  \\
2&
$-{\cal C}^{\gamma N'}_{zz'}$&
$-{\cal C}^{N' V}_{z'x'}$&
$ {\cal C}^{\gamma N' V}_{zz'y'}$&
$-{\cal C}^{N' V}_{z'z'}$&
$-{\cal C}^{N' V}_{z'4}$&
$ {\cal C}^{\gamma N' V}_{zz'5}$
 \\
&&&&&&  \\
\tableline
&&&&&&  \\
3&
$-{\cal C}^{\gamma N N'}_{zxy'}$&
$-{\cal C}^{N N' V}_{xy'x'} $  &
$ -{\cal C}^{\gamma N  V}_{yzy'}$&
$-{\cal C}^{N N' V}_{xy'z'} $ &
$-{\cal C}^{N N' V}_{xy'4} $ &
$ -{\cal C}^{\gamma N  V}_{xyy'} $
 \\
&&&&&&  \\
&
$ $&
$  {\cal C}^{N N' V}_{xz5}$  &
 &
$  -{\cal C}^{N N' V}_{xz4}$ &
$  {\cal C}^{N N' V}_{xzz'}$ &
$  -{\cal C}^{\gamma N  V}_{yz5}$
 \\
&&&&&&  \\
\tableline
&&&&&&  \\
4&
$-{\cal C}^{N N'}_{yy'}$&
${\cal C}^{\gamma V}_{y6}$&
${\cal C}^{\gamma V}_{xy'} $&
${\cal C}^{\gamma V}_{x4}$&
${\cal C}^{\gamma V}_{yz'}$&
${\cal C}^{\gamma V}_{x5} $
 \\
&&&&&&  \\
&
 &
 &
$  -{\cal C}^{N N' V}_{yy'y'}$&
 &
 &
$  -{\cal C}^{N N' V}_{yy'5}$
 \\
&&&&&&  \\
\tableline
&&&&&&  \\
5&
$-{\cal C}^{\gamma N N'}_{zzy'}$&
${\cal C}^{\gamma N  V}_{xx6}$&
$-{\cal C}^{\gamma N  V}_{yxy'}$&
$ -{\cal C}^{\gamma N  V}_{xx4}$ &
${\cal C}^{\gamma N  V}_{xxz'} $&
${\cal C}^{\gamma N  V}_{yx5}$
 \\
&&&&&&  \\
&
 &
${\cal C}^{N N' V}_{zy'x'} $ &
 &
${\cal C}^{N N' V}_{zy'z'}$&
${\cal C}^{N N' V}_{zy'4}  $&

 \\
&&&&&&  \\
\tableline
&&&&&&  \\
6&
$-{\cal C}^{N N'}_{xx'}$&
&
${\cal C}^{N N' V}_{xx'y'}$&
&
&
${\cal C}^{N N' V}_{xx'5}$
 \\
&&&&&&  \\
\tableline
&&&&&&  \\
7&
${\cal C}^{\gamma N N'}_{zyx'}$&
$ {\cal C}^{\gamma N' V}_{xz'6}$ &
$- {\cal C}^{\gamma N' V}_{yz'y'}$&
$-{\cal C}^{\gamma N' V}_{xz'4} $ &
$  {\cal C}^{\gamma N' V}_{xz'z'}$ &
$- {\cal C}^{\gamma N' V}_{yz'5}$
 \\
&&&&&&  \\
&
 &
$ {\cal C}^{N N' V}_{yx'x'}$ &
 &
${\cal C}^{N N' V}_{yx'z'}  $ &
${\cal C}^{N N' V}_{yx'4} $ &

 \\
&&&&&&  \\
\tableline
 &&&&&&  \\
8&
${\cal C}^{N N'}_{zx'}$&
&
$-{\cal C}^{N N' V}_{zx'y'}$&
&
&
$-{\cal C}^{N N' V}_{zx'5}$ \\
 &&&&&&
\end{tabular}\newpage
\begin{tabular}{ccccccc}
$\alpha\backslash\beta$&1&2&3&4&5&6  \\
&&&&&&  \\
\tableline
&&&&&&  \\
9&
${\cal C}^{\gamma N}_{zz}$&
$-{\cal C}^{N V}_{zx'}$&
$ -{\cal C}^{\gamma N  V}_{zzy'}$&
$-{\cal C}^{N V}_{zz'}$&
$-{\cal C}^{N V}_{z4}$&
$ -{\cal C}^{\gamma N  V}_{zz5}$
 \\
&&&&&&  \\
\tableline
&&&&&&  \\
10&
$-T$&
  &
${\cal C}^{N' V}_{yy} $ &
&
 &
$ {\cal C}^{N' V}_{y5} $
  \\
&&&&&&  \\
&
 &
&
$ -{\cal C}^{N V}_{yy'}$&
&
&
$ -{\cal C}^{N V}_{y5}$
 \\
&&&&&&  \\
&
 &
$ {\cal C}^{\gamma N' V}_{yy'6} $  &
$  {\cal C}^{\gamma N' V}_{xy'y'}$ &
 $- {\cal C}^{\gamma N' V}_{yy'4} $ &
$ {\cal C}^{\gamma N' V}_{yy'z'}$  &
$   {\cal C}^{\gamma N' V}_{xy'5}$
  \\
&&&&&&  \\
&
 &
$  -{\cal C}^{\gamma N  V}_{zyx'}$  &
  &
$  -{\cal C}^{\gamma N  V}_{zyz'}$ &
  &
$ -{\cal C}^{\gamma N  V}_{zy6}$
  \\

&&&&&&  \\
\tableline
&&&&&&  \\
11&
${\cal C}^{\gamma N}_{zx}$&
$-{\cal C}^{N V}_{xx'}$&
$ -{\cal C}^{\gamma N  V}_{zxy'}$&
$-{\cal C}^{N V}_{xz'}$&
$-{\cal C}^{N V}_{x4}$&
$ -{\cal C}^{\gamma N  V}_{zx5}$
 \\
&&&&&&  \\
\tableline
&&&&&&  \\
12&
$P_{N'}$&
$ {\cal C}^{\gamma N' V}_{zy'x'} $&
$ -{\cal C}^{\gamma N  V}_{xyy'}$ &
$ {\cal C}^{\gamma N' V}_{zy'z'} $&
$ {\cal C}^{\gamma N' V}_{zy'4} $&
$ -{\cal C}^{\gamma N  V}_{xy5}$ \\
&&&&&&  \\
&
 &
$  -{\cal C}^{\gamma N  V}_{yy6}$&
$  {\cal C}^{ N'  V}_{y'y'}$ &
$   {\cal C}^{\gamma N  V}_{yy4}$&
$   -{\cal C}^{\gamma N  V}_{yyz'}$&
 $  {\cal C}^{ N'  V}_{y'5}$ \\
&&&&&&  \\
\tableline
&&&&&&  \\
13&
${\cal C}^{N N'}_{xz'}$&
&
${\cal C}^{N N' V}_{xz'y'}$ &
&
&
${\cal C}^{N N' V}_{xz'5}$
 \\
&&&&&&  \\
\tableline
&&&&&&  \\
14&
${\cal C}^{\gamma N N'}_{zyz'}$&
$ -{\cal C}^{\gamma N' V}_{xx'6}  $  &
${\cal C}^{\gamma N' V}_{yx'y'}$&
$ {\cal C}^{\gamma N' V}_{xx'4} $  &
$  -{\cal C}^{\gamma N' V}_{xx'z'} $   &
${\cal C}^{\gamma N' V}_{yx'y'}$
 \\
&&&&&&  \\
&
 &
${\cal C}^{N N' V}_{yz'x'}  $  &
 &
$ {\cal C}^{N N' V}_{yz'z'}   $  &
$ {\cal C}^{N N' V}_{yz'4} $   &

 \\
&&&&&&  \\
\tableline
&&&&&&  \\
15&
${\cal C}^{N N'}_{zz'}$&
&
$-{\cal C}^{N N' V}_{zz'y'}$&
&
&
$-{\cal C}^{N N' V}_{zz'5}$ \\
&&&&&&  \\
\tableline
&&&&&&  \\
16&
$-{\cal C}^{\gamma N'}_{zx'}$&
$-{\cal C}^{N' V}_{x'x'}$&
$- {\cal C}^{\gamma N' V}_{zx'y'}$&
$-{\cal C}^{N' V}_{x'z'}$&
$-{\cal C}^{N' V}_{x'4}$&
$- {\cal C}^{\gamma N' V}_{zx'5}$
  \\
&&&&&&
\end{tabular}
 \label{matrixtoobs1}
 \end{table}

\mediumtext

\begin{table}
\caption{ Spin Observables in BHP form as products of $\Gamma^\alpha,$ and
 $\omega^\beta $ matrices. Here the $\alpha=1 \cdots 16,
  \beta= 1-6 $ range is displayed.}
\begin{tabular}{ccccccc}
$\alpha\backslash\beta$ &7&8&9&A&B&C \\
\tableline
&&&&&&  \\
1&
${\cal C}^{\gamma V}_{z 6}$&
$T_{22}$&
$T_{20}$&
$T_{y' y'} $&
&
\\
&&&&&&  \\
&
 &
 &
 &
$  -{\cal C}^{\gamma N  N'}_{xyy'}$&
&
\\
&&&&&&  \\
\tableline
&&&&&&  \\
2&
$-{\cal C}^{N' V}_{z' 6}$&
$ {\cal C}^{\gamma N' V}_{z z' 7}$&
$ {\cal C}^{\gamma N' V}_{z z '8}$&
$-{\cal C}^{\gamma N N'}_{y y x'}$&
&
\\
&&&&&&  \\
\tableline
&&&&&&  \\
3&
$-{\cal C}^{N N' V}_{x z x'}$ &
&
&
${\cal C}^{\gamma N}_{y z}$&
$ -{\cal C}^{\gamma N  V}_{y z 7}$&
$ -{\cal C}^{\gamma N  V}_{y z 8}$
\\
&&&&&&  \\
\tableline
&&&&&&  \\
4&
$-{\cal C}^{\gamma V}_{y x'}$&
$  -{\cal C}^{N N' V}_{y y' 7}$&
 $ -{\cal C}^{N N' V}_{y y '8}$&
$\Sigma$&
${\cal C}^{\gamma V}_{x 7}$&
${\cal C}^{\gamma V}_{x 8}$
\\
&&&&&&  \\
\tableline
&&&&&&  \\
5&
${\cal C}^{N N' V}_{z y' 6} $&
&
&
${\cal C}^{\gamma N}_{y x}$&
${\cal C}^{\gamma N  V}_{y x 7}$&
${\cal C}^{\gamma N  V}_{y x 8}$
\\
&&&&&&  \\
&
$  -{\cal C}^{\gamma N  V}_{xxx'}$&
&
&
 &
 &

\\
&&&&&&  \\
\tableline
&&&&&&  \\
6&
&
${\cal C}^{N N' V}_{x x' 7}$&
${\cal C}^{N N' V}_{x x' 7}$&
$ -{\cal C}^{\gamma N N' }_{xzz'}$&
&
 \\
&&&&&&  \\
\tableline
&&&&&&  \\
7&
$   -{\cal C}^{\gamma N' V}_{x z' x'}$ &
&
&
$-{\cal C}^{\gamma N'}_{y z'}$&
$- {\cal C}^{\gamma N' V}_{y z' 7}$&
$- {\cal C}^{\gamma N' V}_{y z' 8}$
\\
&&&&&&  \\
&
${\cal C}^{N N' V}_{y x' 6} $ &
&
 &
&
 &

\\
&&&&&&  \\
\tableline
&&&&&&  \\
8&
&
 &
$  -{\cal C}^{\gamma N  N'}_{x x z'}$&
&
&
\\
&&&&&&  \\
&
&
$-{\cal C}^{N N' V}_{z x' 7}$&
$-{\cal C}^{N N' V}_{z x' 8} $&
&
&
\\
&&&&&&
\end{tabular}\newpage
\begin{tabular}{ccccccc}
$\alpha\backslash\beta$ &7&8&9&A&B&C \\
\tableline
&&&&&&  \\
9&
$-{\cal C}^{N V}_{z6}$&
$ -{\cal C}^{\gamma N  V}_{zz7}$&
$ -{\cal C}^{\gamma N  V}_{zz8}$&
$-{\cal C}^{\gamma N N'}_{yxy'}$&
&
\\
&&&&&&  \\
\tableline
&&&&&&  \\
10&
$- {\cal C}^{\gamma N' V}_{yy'x'}$ &
${\cal C}^{N' V}_{y7} $&
${\cal C}^{N' V}_{y8} $&
${\cal C}^{\gamma N'}_{xy'}$&
 ${\cal C}^{\gamma N' V}_{xy'7}$&
 ${\cal C}^{\gamma N' V}_{xy'8}$
\\

&&&&&&  \\
&
&
$ -{\cal C}^{N V}_{y7}$&
$ -{\cal C}^{N V}_{y8}$&
&
&
\\
&&&&&&  \\
\tableline
&&&&&&  \\
11&
$-{\cal C}^{N V}_{x6}$&
$ -{\cal C}^{\gamma N  V}_{zx7}$&
$ -{\cal C}^{\gamma N  V}_{zx8}$&
$-{\cal C}^{\gamma N N'}_{yzy'}$&
&
\\
&&&&&&  \\
\tableline
&&&&&&  \\
12&
$ {\cal C}^{\gamma N' V}_{zy'6} $&
$ {\cal C}^{  N' V}_{y'7} $&
$ {\cal C}^{  N' V}_{y'8} $&
$-{\cal C}^{\gamma N }_{xy}$   &
$ -{\cal C}^{\gamma N  V}_{xy7}$&
$ -{\cal C}^{\gamma N  V}_{xy8}$\\
&&&&&&  \\
&&&&&&  \\
&
$   {\cal C}^{\gamma N  V}_{yyx'}$&
&
&
    &
 &
 \\
&&&&&&  \\
\tableline
&&&&&&  \\
13&
&
$-{\cal C}^{N N' V}_{xz'7}$ &
$-{\cal C}^{N N' V}_{xz'8}$ &
$ {\cal C}^{\gamma N  N'}_{xzx'}$&
&
\\
&&&&&&  \\
\tableline
&&&&&&  \\
14&
$   {\cal C}^{\gamma N' V}_{xx'x'} $ &
&
&
${\cal C}^{\gamma N'}_{yx'}$&
${\cal C}^{\gamma N' V}_{yx'7}$&
${\cal C}^{\gamma N' V}_{yx'8}$\\
&&&&&&  \\
&
${\cal C}^{N N' V}_{yz'6} $ &
&
&
 &
 &
 \\
&&&&&&  \\
\tableline

&&&&&&  \\
15&
&
$-{\cal C}^{N N' V}_{zz'7}$&
$-{\cal C}^{N N' V}_{zz'8}  $&
&
&
\\
&&&&&&  \\
&
&
 &
$  {\cal C}^{\gamma N  N'}_{xxx'}$&
&
&
\\
&&&&&&  \\
\tableline
&&&&&&  \\
16&
$-{\cal C}^{N' V}_{x'6}$&
$- {\cal C}^{\gamma N' V}_{zx'7}$&
$- {\cal C}^{\gamma N' V}_{zx'8}$&
${\cal C}^{\gamma N N'}_{yyz'}$&
&
 \\
&&&&&&  \\
 \end{tabular}
 \label{matrixtoobs2}
 \end{table}

\end{document}